\documentclass[journal]{IEEEtran}
\usepackage{cite}

\usepackage{graphicx}
\ifCLASSINFOpdf
\graphicspath{{.}}
\else
\fi
\usepackage[cmex10]{amsmath}
\usepackage{algorithm}
\usepackage[tight,footnotesize]{subfigure}
\usepackage{url}

\begin{document}
%
\title{A multiscale Laplacian of Gaussian (LoG) filtering approach to pulmonary nodule detection from whole-lung CT scans}
%
%
%

        
\author{Sergei~V.~Fotin,
        David~F.~Yankelevitz,
        Claudia~I.~Henschke,
        and~Anthony~P.~Reeves}
\maketitle

\begin{abstract}
Candidate generation, the first stage for most computer aided detection (CAD) systems, rapidly scans the entire image data for any possible abnormality locations, while the subsequent stages of the CAD system refine the candidates list to determine the most probable or significant of these candidates. The candidate generator creates a list of the locations and provides a size estimate for each candidate. A multiscale scale-normalized Laplacian of Gaussian (LoG) filtering method for detecting pulmonary nodules in whole-lung CT scans, presented in this paper, achieves a high sensitivity for both solid and nonsolid pulmonary nodules. 

The pulmonary nodule LoG filtering method was validated on a size-enriched database of 706 whole-lung low-dose CT scans containing 499 solid ($\ge$~4 mm) and 107 nonsolid ($\ge$~6 mm) pulmonary nodules. The method achieved a sensitivity of 0.998 (498/499) for solid nodules and a sensitivity of 1.000 (107/107) for nonsolid nodules. Furthermore, compared to radiologist measurements, the method provided low average nodule size estimation error of 0.12~mm for solid and 1.27~mm for nonsolid nodules. The average distance between automatically and manually determined nodule centroids were 1.41~mm and 1.43~mm, respectively.
\end{abstract}


%
\IEEEpeerreviewmaketitle

\section{Introduction}
%
%
%
%
\IEEEPARstart{L}{ung} cancer is the leading cause of cancer related death in the world. It is estimated that more than 160000 people die from it every year in the United States\footnote[1]{American Cancer Society. Cancer Facts and Figures, 2012, \url{http://www.cancer.org/}, accessed February 21, 2012.}. Computed tomography of the chest is one of the imaging modalities that can be used to diagnose lung cancer. Regular clinical practice involves manual visual inspection of hundreds cross-sectional slices of a patient's CT scan for small pulmonary nodules that can manifest early lung cancer. However, radiologists routinely miss nodules~\cite{wormanns2005detection, leader2005pulmonary} due to fatigue and the error-prone nature of the work, which may ultimately lead to incorrect diagnostic decisions. It has been shown that sensitivity of detection can be improved significantly by introducing a computer algorithm that reviews the same CT image a second time~\cite{KMCE05, MDGM06, FBLZ07} and detects nodules the human reader may have missed. Alternatively, a computer algorithm may work independently and present the detection results to an operator. In each of these uses high sensitivity computer-aided detection (CAD) systems are necessary.

There have been a number of CAD systems for pulmonary nodules in CT scans reported in literature; for survey see the review papers by Sluimer et al.~\cite{ISAS06}, Li~\cite{QLQL07}, Chan et al.~\cite{HCLH08}. 

Traditionally CAD systems are considered to consist of two primary components: a candidate generator (CG) followed by false positive eliminator (FPE). The main driving philosophy for this design is computational efficiency; that is, the candidate generator can rapidly eliminate most of the image space from consideration so that the false positive eliminator can apply more computational resources to the analysis of these selected regions. In most systems, the clinical performance (sensitivity/number of false positives) of a CAD system is determined by jointly optimizing the combined performance of these two stages. In this work, we propose that the development of these two components may be conducted independently and we provide a characterization and implementation for a clinically effective candidate generator.

We define the functions and requirements of these two stages as follows. The prime objective of the candidate generator is to reduce the nodule search space without missing any nodules. If such a generator is inefficient (provides too many false positives) then this will increase the time required for the false positive elimination stage. However, if the candidate generator misses a single nodule then this can never be recovered by the false positive stage and the of the whole CAD system is compromised since the system sensitivity will be strictly limited. The primary function of the false positive elimination stage is to discard suspicious regions when they do not actually contain any pulmonary nodules. The primary design goal should be to remove all false positives irrespective of the computational cost involved.

Given development of the above two systems there is a practical consideration that there be a limit on the computation resources/time for a computer algorithm to apply to each case. Once the CG and FPE have been independently optimized for their specific function the appropriate time optimization can then be made to reduce the computation time to the provided limits. That is, one can improve execution time by either reducing the sensitivity of the CG providing less candidates to analyze or by reducing the specificity of the FPE by simplifying the FPE algorithm. Note under this specification the CG provides for one means of computation time reduction and that, given unlimited time, the FPE could be used without the CG to evaluate all possible nodule locations and sizes to provide the same outcome but requiring vastly more computation time.

One advantage of the above scheme is to provide for the independent optimization of the CAD components and to use the practical time constraint as the real limitation on ideal performance giving motivation to consider implementation efficiency as the final stage of system optimization. That is, we avoid the traditional comingling of clinical efficacy and computational efficiency by considering these items separately and first focusing on the primary objective of the former.

The candidate generator traditionally reduces the search space for FPE by specifying the image regions or specific locations in 3D image space that may contain nodules. That is it identifies, approximate 3D location and size for candidate nodules.  Minimizing the uncertainty in the estimates for these four nodule parameters is an important CG design consideration, since more precise specification further constrains the search space for the FPE. Therefore, we consider the precision of location and size to be important design characteristics for the CG in addition to the primary characteristic of minimum number of false positives given close to 100\% sensitivity.

The prime characteristic for the candidate generator is to not miss a single nodule; i.e. to have sensitivity as close to 100\% as possible. Such a generator will typically provide far more candidates than have been traditionally considered in the literature although they still reduce the 4D (x,y,z and scale) search space by many orders of magnitude. Once we have such a CG it is very simple to design additional filters that will reduce the number of candidates to the level of traditional CG (doing so implies that the FPE is computationally time limited given the higher candidate load). The design challenge then becomes to improve these filters such that any compromise in sensitivity is minimized.

Example algorithms for nodule candidate generation and corresponding sensitivities evaluated on datasets containing at least 100 pulmonary nodules can be found in literature~\cite{zhao2003automatic, AFAE04, SAAR05, AEAR07, JPBZ08, QLFL08, murphy2009large, messay2010new} and are summarized in Table~\ref{tab:perfcomparison}. Reported sensitivities in these studies ranged from 88.3\% to 98.7\%. 

\begin{table*}[!t]
\renewcommand{\arraystretch}{1.3}
\caption{Examples of reported candidate generators with their initial sensitivities.}
\label{tab:perfcomparison}
\begin{center}
\small
\begin{tabular}{|l|l|l|l|l|}
\hline
First author & Algorithm	& Nodules	& Sensitivity & False positive rate\\
\hline
Zhao, 2003 &	Local density maximum	& 266, $\ge$ 2 mm & 0.944 & 906\\
Farag, 2004	& Template matching &	130 & 0.846 & 49 (per slice)\\
Armato, 2005 &	Multiple gray-level thresholding &	470 & 0.883 & 539 \\
Enquobahrie, 2007 &	Binary image features	& 499, $\ge$ 4 mm & 0.958 & 3785\\
Pu, 2008 & Signed distance field	& 184, $\ge$ 4 mm & 0.951 & 1200\\
Li, 2008 & Selective enhancement filters &	153, $\ge$ 4 mm & 0.987 & 140\\
Murphy, 2009 (3 datasets) &	Local image features & 1525/1688/768, $\ge$ 3 mm &	0.972/0.977/0.982 & 649.0/750.5/752.1\\
Messay, 2010 & Multiple gray-level thresholding &	143, $\ge$ 3 mm & 0.923 & 900-1200\\
\hline
\end{tabular}
\end{center}
\end{table*}

Zhao~\cite{zhao2003automatic} used the multiple-level thresholding that involves segmenting the lungs, setting an image intensity threshold at a certain value and detecting connected three-dimensional objects. The threshold is then decreased in a step-wise manner recovering less-dense objects and extending the objects detected at higher thresholds. The decision whether to treat a connected object as a nodule candidate is based on a set of rules that incorporate size, density, and a variable describing the acceptable change of an object's volume. There are additional parameters that specify intensity threshold bound and step size. Another multiple intensity level thresholding technique was used by Armato et al.~\cite{SAAR05}. Here, instead of decreasing threshold, the authors increased it stepwise and recorded all the objects smaller than a certain size. The algorithm described by Farag et al.~\cite{AFAE04} involves template matching of the local lung regions with predefined nodule templates of various size and shape using normalized cross-correlation measure. Since the dimensionality of the search space was high, the authors did not perform an exhaustive search, but instead, resorted to a genetic algorithm, that provided the position and initial size of the candidate templates. Pu et al.~\cite{JPBZ08} first obtained a binary image of a CT scan using an empirically selected threshold. Nodule candidate generation was quite simple --- the authors calculated signed Euclidean distance fields and recorded local maxima and corresponding distances as nodule candidates. The work of Li et al.~\cite{QLFL08} relies on the use of filters for nodule enhancement and suppression of normal anatomic structures such as blood vessels and lung walls. These filters are multiscale in nature and are able to characterize the local image structures such as sphere, cylinder or plane. This discrimination is possible due to computing local image curvature using second-order image derivatives. After the original CT scan is filtered, thresholding identifies nodule candidates. In the work of Murphy et al.~\cite{murphy2009large}, nodule candidates were identified by computing shape index and curvedness for each image location. Voxels having these values above a certain threshold were clustered together and adjusted to form nodule candidates. Messay et al.~\cite{messay2010new} paired intensity thresholding and morphological opening to obtain the candidate mask at multiple threshold levels. The final candidate mask was constructed as a union of these masks. The method described by Enquobahrie et al.~\cite{AEAR07} is capable of identifying solid nodules and consists of two modules: for detecting isolated (inside lung parenchyma attached to small pulmonary vessels and airways) and attached (adjacent to lung wall, mediastinum surface) nodules. Briefly, the first module works with the thresholded image of the lungs and looks for pixel clusters that have a certain minimum size and limited extent. The second module is based on morphological processing and analysis of the lung boundary shape for identifying protrusions of significant size.  Both modules have an associated set of parameters that need to be tuned to achieve maximum performance. Since this method was previously developed and used in our research group, it will be used as a reference for the experiments described further in this paper.

All of the candidate generation techniques mentioned above have either thresholds or parameters that need to be fine-tuned to achieve the best performance. Even though such flexibility may result in a very high sensitivity, optimization of these generators to datasets with different image acquisition parameters or target nodule size ranges may require additional work. In order to minimize the effect of these issues, in the design of our nodule candidate generator the number of control parameters was kept to a necessary minimum. In addition, the majority of previously developed candidate generators rely on correct segmentation of the lung volume that is critical for proper localization of nodules adjacent to lung wall. In contrast, our candidate generation scheme is the same for all candidates and does not depend on the lung segmentation outcome.

The candidate generator presented in this paper is based on multiscale Laplacian of Gaussian (LoG) filtering. LoG filtering has been used for multiscale analysis in many computer vision and image analysis applications. Use of LoG filtering for detecting edges on digital images was first proposed by Marr and Hildreth~\cite{marr1980theory}. Subsequently, LoG filtering was used for enhancing image "blobs" for locating the aorta in MR imagery by Jiang and Merickel~\cite{jiang1989identification}; and for identifying image texture elements by Blostein and Ahuja~\cite{blostein2002shape}. The method of detecting "blobs" as scale-space structures was formalized by Lindeberg~\cite{TLTL93} and later employed by many authors. LoG-based "blob" detection has been used for nodule detection in two-dimensional chest radiographs by Schilham et al.~\cite{ASBG03}. The application of LoG filtering for localization of nodules in the three-dimensional CT scans was first proposed by Reeves et al.~\cite{ARAC06} and evaluated with respect to the nodule size estimation properties in the works of Jirapatnakul et al.~\cite{AJSF09} and Diciotti et al.~\cite{diciotti2010log}. The previous application of LoG filtering with respect to computerized nodule detection from CT has been limited to the work of Dolejsi and Kybic~\cite{MDJK07}.

\section{Method}

The presented method takes into account the distinction between different nodule types. In the following section two fundamentally different classes of nodules are described and a specific approach tailored to each nodule type is then developed.

\subsection{Nodule types dichotomy}

A solid pulmonary nodule is an approximately spherical lesion having an image intensity similar to that of soft tissue (median = \mbox{-294 HU}, standard deviation = 164~HU as found by Browder~\cite{WBWB07}). This compares to the intensity of air which, by definition, has a value of \mbox{-1000~HU}. The size range for pulmonary nodules is from 3 mm to 3 cm in diameter~\cite{austin1996glossary}. The majority of the image intensity in the lungs is the aerated lung parenchyma which has an image intensity approaching that of air (median = \mbox{-810 HU}, stdev = 63 HU~\cite{WBWB07}). Other normal visible structures in the lungs are airways and blood vessels. These are branching cylindrical structures that have similar image intensity to soft tissue.  The airways also have a lumen that is the intensity of air. The minimum solid nodule diameter to consider radiological lung examination non-negative, varies in the range of 4.5 - 5 mm~\cite{IELCAP, DXHG06, lopes2009design}.

The nonsolid nodule is a second class of pulmonary nodule that, unlike the solid nodule, is caused by layers of cells lining the alveoli and the airways. In overall appearance it has a density that is slightly more than that of the lung parenchyma since there is still a significant amount of air within the airways and alveoli (median = \mbox{-680 HU}, standard deviation = 58 HU~\cite{WBWB07}). In these lesions, more dense vessels and larger airways may be visible. In general, these lesions are also approximately spherical. Nonsolid nodules are harder to see compared to solid nodules and when small in size are difficult to discriminate from other parenchyma density variations. For this reason clinical significance is usually considered for nonsolid nodules having a minimum size in the range 8 - 10 mm.

\begin{figure}
\centering
\subfigure[]{\label{fig:nodsolid}\includegraphics[width=0.95\linewidth]{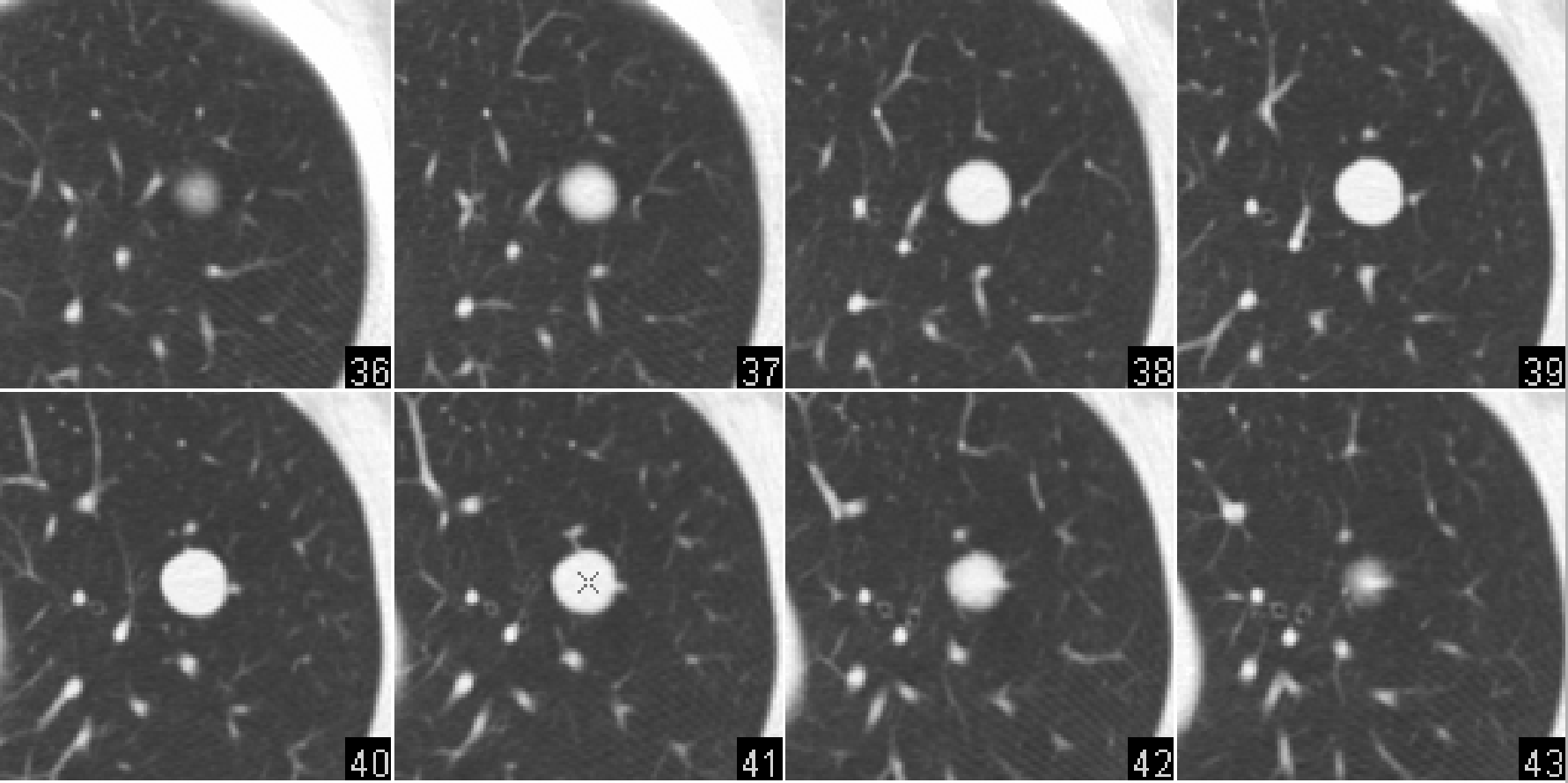}}\\
\subfigure[]{\label{fig:nodnonsolid}\includegraphics[width=0.95\linewidth]{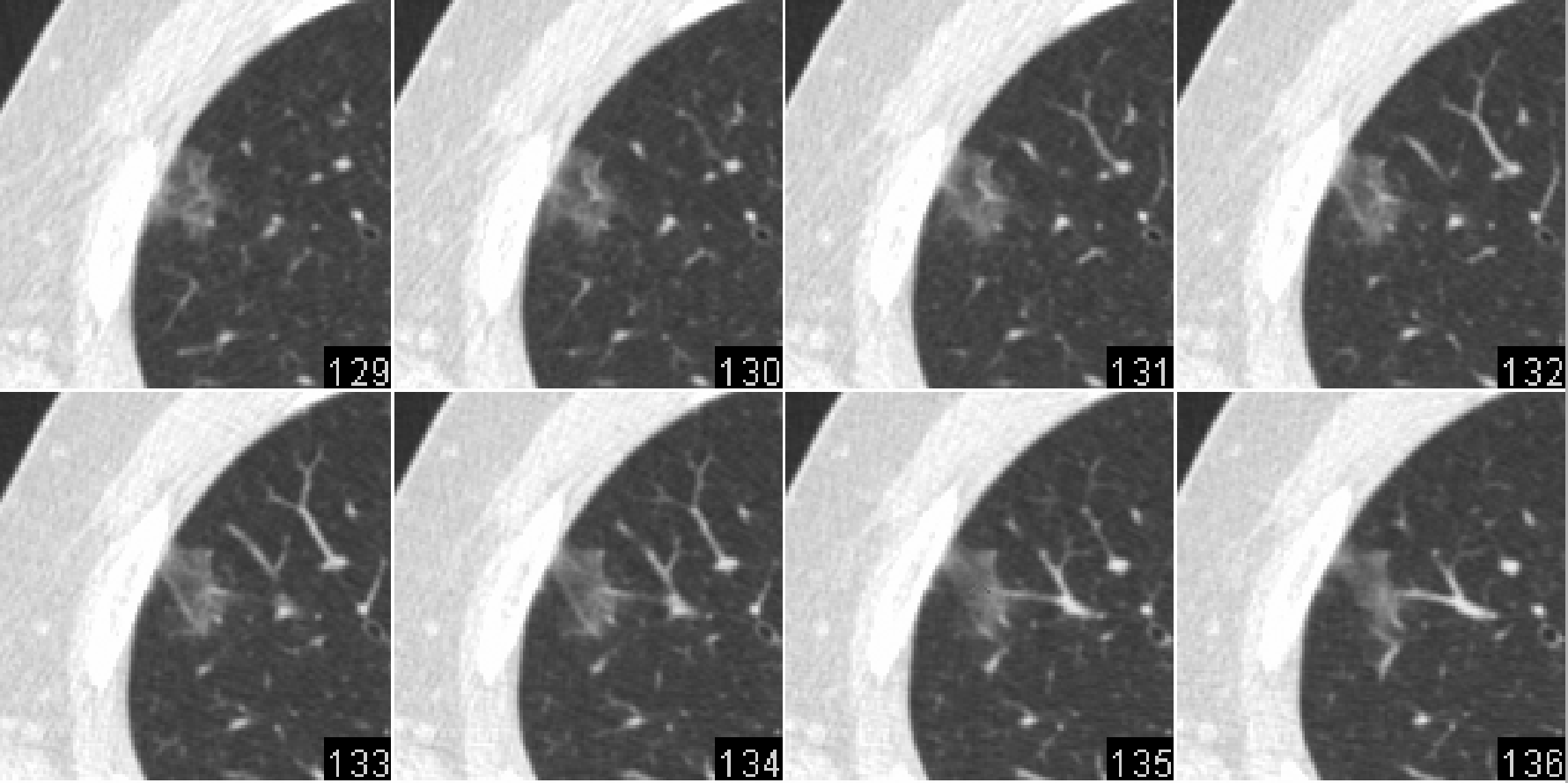}}
\caption{Appearance of a solid (a) and nonsolid (b) pulmonary nodule.}
\label{fig:solidnonsolid}
\end{figure}

Typical pulmonary nodules of both types are shown in Figure~\ref{fig:solidnonsolid}. Nodules may have variation in size, shape, intensity profile and attachment morphology.

In the following section the model of a solid nodule and the corresponding detection method are presented. The solid nodule model is then extended to deal with the more complex case of nonsolid nodules. 

\subsection{Solid pulmonary nodule image model}

The objective of the candidate generator is to identify pulmonary nodules from other normal structures in the lung region that have a density greater than the lung parenchyma background. For this reason, a solid pulmonary nodule is modelled as a solid spherical structure with the intensity of soft tissue (equal to 1) on a background having the same intensity as lung parenchyma (equal to 0).  First, the properties of such a model are explored in isolation from other objects. Then, the applicability of the model is investigated in the presence of interference from the structures resembling pulmonary vessels and chest walls. Finally, the the method is tested on real dataset where nodules are not perfectly spherical and may have more complex attachments.

The Laplacian of Gaussian has been described as a detector that responds to "bright regions on dark background or vice versa"~\cite{TLTL93}. That is the general definition that is used for the term "blob" in this paper. Marr in his original work~\cite{marr1976early, marr1980theory} more formally defined a "blob" as a primitive compact image element enclosed in the closed contour made of the LoG zero-crossings.

\subsection{Blob detection as scale-space normalized LoG filtering}

A method that detects the location and estimates the scale of blob-like structures is described by Lindeberg~\cite{TLTL98}. Identification of "blobs" is accomplished by finding scale-space maxima and minima of scale-normalized Laplacian
\begin{equation}
\nabla_{norm}^2L(X,\sigma):\Re^3 \times \Re \to \Re.
\end{equation}
Here $L(X,\sigma)$  is the scale-space representation obtained for image $I(X)$ by convolving it with Gaussian kernel $G(X,\sigma)$ at continuous set of scales:
\begin{equation}
L(X,\sigma) = G(X,\sigma)*I(X).
\end{equation}
In this paper, the following notation is used for a three-dimensional coordinate vector: $X = (x, y, z)$; and the standard deviation of Gaussian kernel $\sigma$ is referred to as the "scale parameter" or, simply, "scale" or "size."

The response function $\nabla_{norm}^2L(X,\sigma)$ may be expressed as follows using the properties of convolution:
\begin{equation}
\nabla_{norm}^2L(X,\sigma) = \nabla_{norm}^2G(X,\sigma)*I(X).
\end{equation}
This operation is effectively the filtering of the original image with scale-normalized LoG kernels of continuously changing scale parameter. 

Normalization is necessary to eliminate the effect of decreasing spatial derivatives with the increase of scale and is defined as the negated multiplication of LoG by $\sigma^2$:
\begin{equation}
\nabla_{norm}^2G(X,\sigma) =  - \sigma ^2\nabla ^2G(X,\sigma).
\end{equation}

By introducing the negative sign for normalization, the computations are brought to the domain of positive real numbers, where "bright" blobs are identified as local maxima of $\nabla_{norm}^2L(X,\sigma)$ instead of local minima. Later in the paper, the negated scale-normalized LoG will be referred to as simply the "normalized LoG." 

The initial set $C$ of nodule candidates is constructed as a subset of local maxima of the response function $\nabla_{norm}^2L(X,\sigma)$, such that their spatial component is located within the lungs:
\begin{equation}
\begin{split}
C = \{&(Y,\sigma):Y \in Lungs, \\
&(Y,\sigma) \in \left\{ \mathop{local max}\;\nabla_{norm}^2L(X,\sigma) \right\} \}.
\end{split}
\end{equation}

\subsection{Properties of the normalized LoG filter}
\begin{figure}
		\centering
		\includegraphics[width=0.3\linewidth]{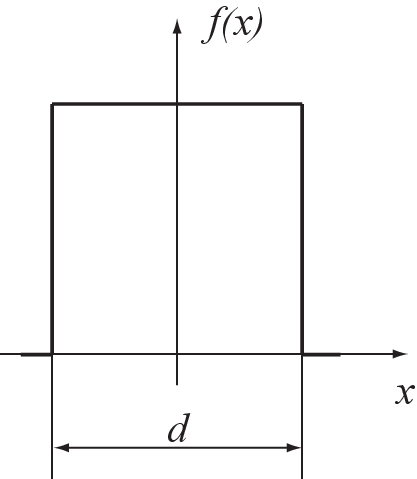}
		\caption{\label{fig:rectangle} Rectangular function as one-dimensional representation of the nodule model.}
\end{figure}

To illustrate the concept of normalized LoG filtering, the nodule model is reduced to a one-dimensional representation --- rectangular function as sketched in Figure~\ref{fig:rectangle}.

\begin{figure*}[t]
		\centering
		\includegraphics[width=0.8\linewidth]{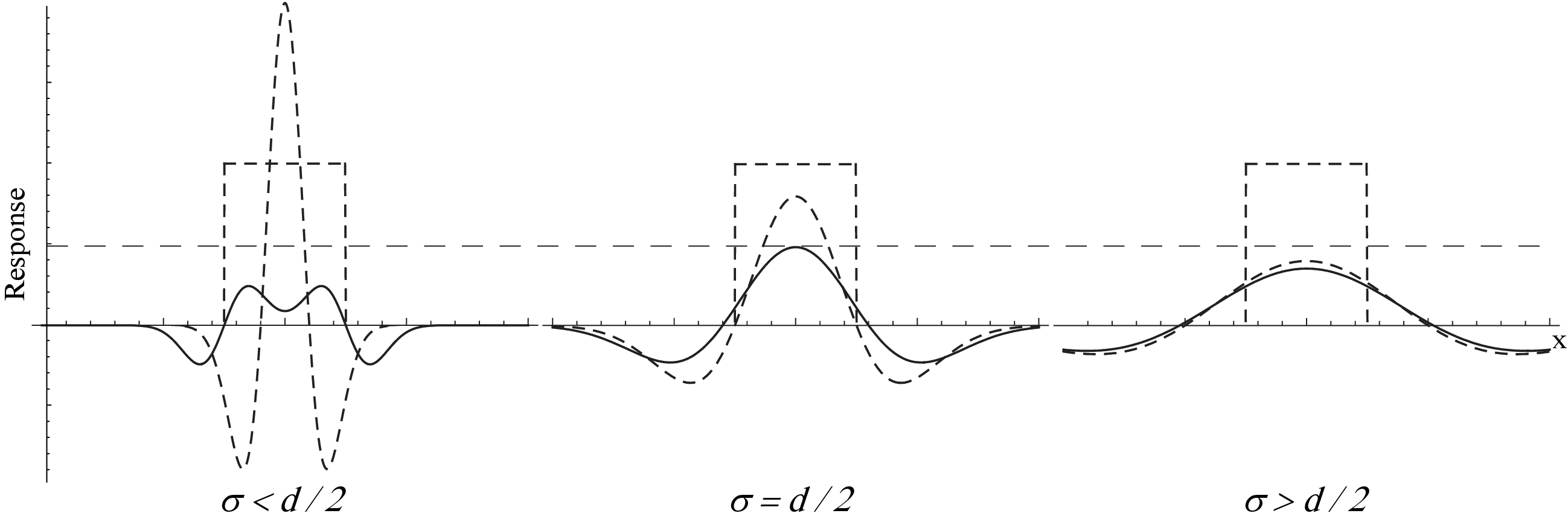}
		\caption{\label{fig:responses1} Illustration of multiscale LoG filtering: normalized LoG kernels of varying scales (curvy dashed line) are convolved with the rectangle function (dashed line). Maximum response (solid line) is observed when the size of the kernel corresponds to the width of the rectangle (i.e. normalized LoG zero-crossings coincide with the "boundary" of the rectangle). Long horizontal dashed line corresponds to the value of maximal response achieved at $\sigma=d/2$.}
\end{figure*}

\begin{figure}
		\centering
		\includegraphics[width=0.9\linewidth]{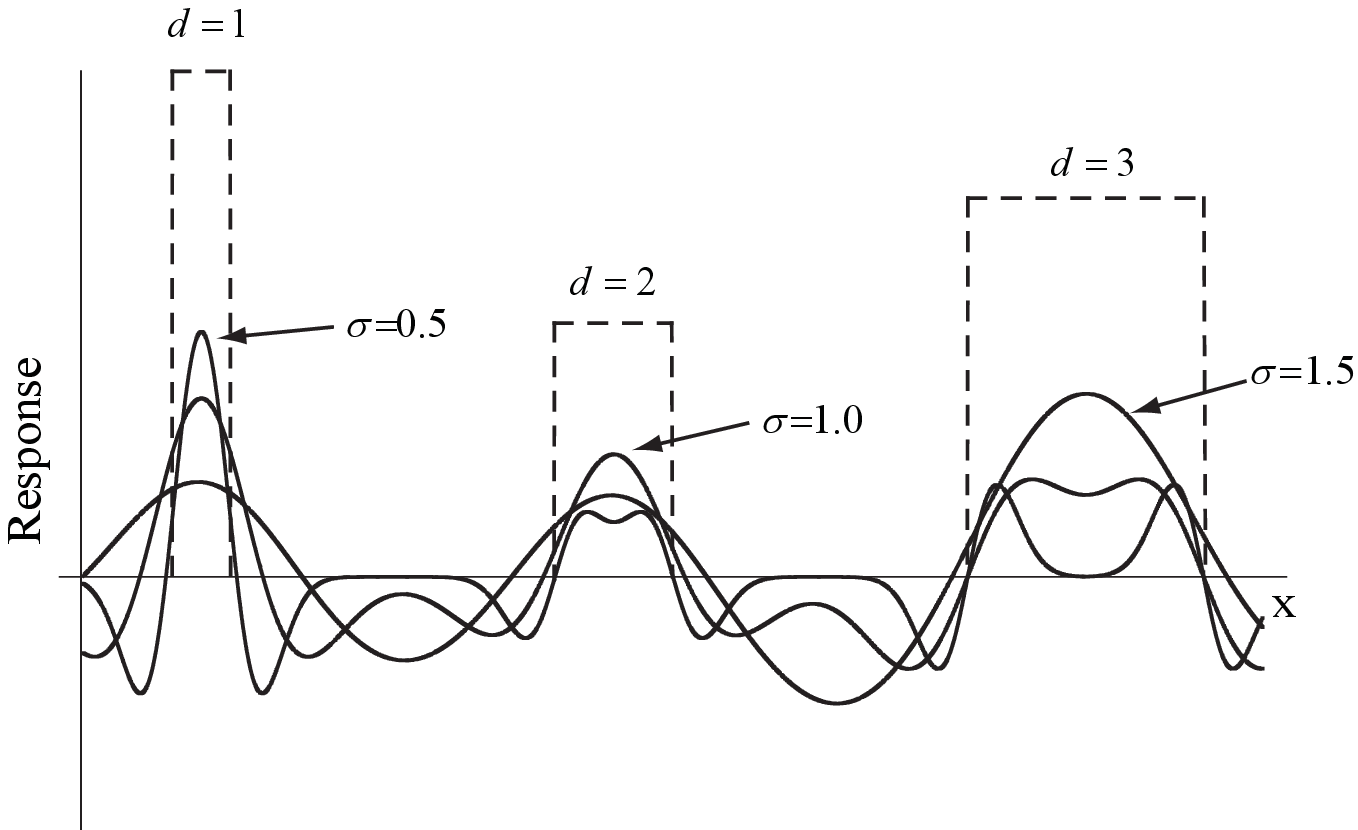}
		\caption{\label{fig:responses2} Illustration of multiscale LoG filtering with respect to three rectangular functions of different size and intensity. Three curves represent responses of the differently sized normalized LoG kernels. Maximum response is proportional to the height of the rectangle and is achieved when the kernel width corresponds to its size.}
\end{figure}

In Figure~\ref{fig:responses1} the responses of the normalized LoG filter to the rectangular function of fixed width are observed. An illustration of the filtering applied to an image having three rectangles of varying widths and intensities is shown in Figure~\ref{fig:responses2}. The maximum response occurs when the LoG kernel is located at the center of the rectangle and when the width of rectangle matches the size of its central lobe: $\sigma = d/2$ (see Appendix~\ref{apx0} for proof). Linearity of the LoG filter causes the response to be proportional to the height of the rectangle.

\begin{figure}
		\centering
		\includegraphics[width=0.75\linewidth]{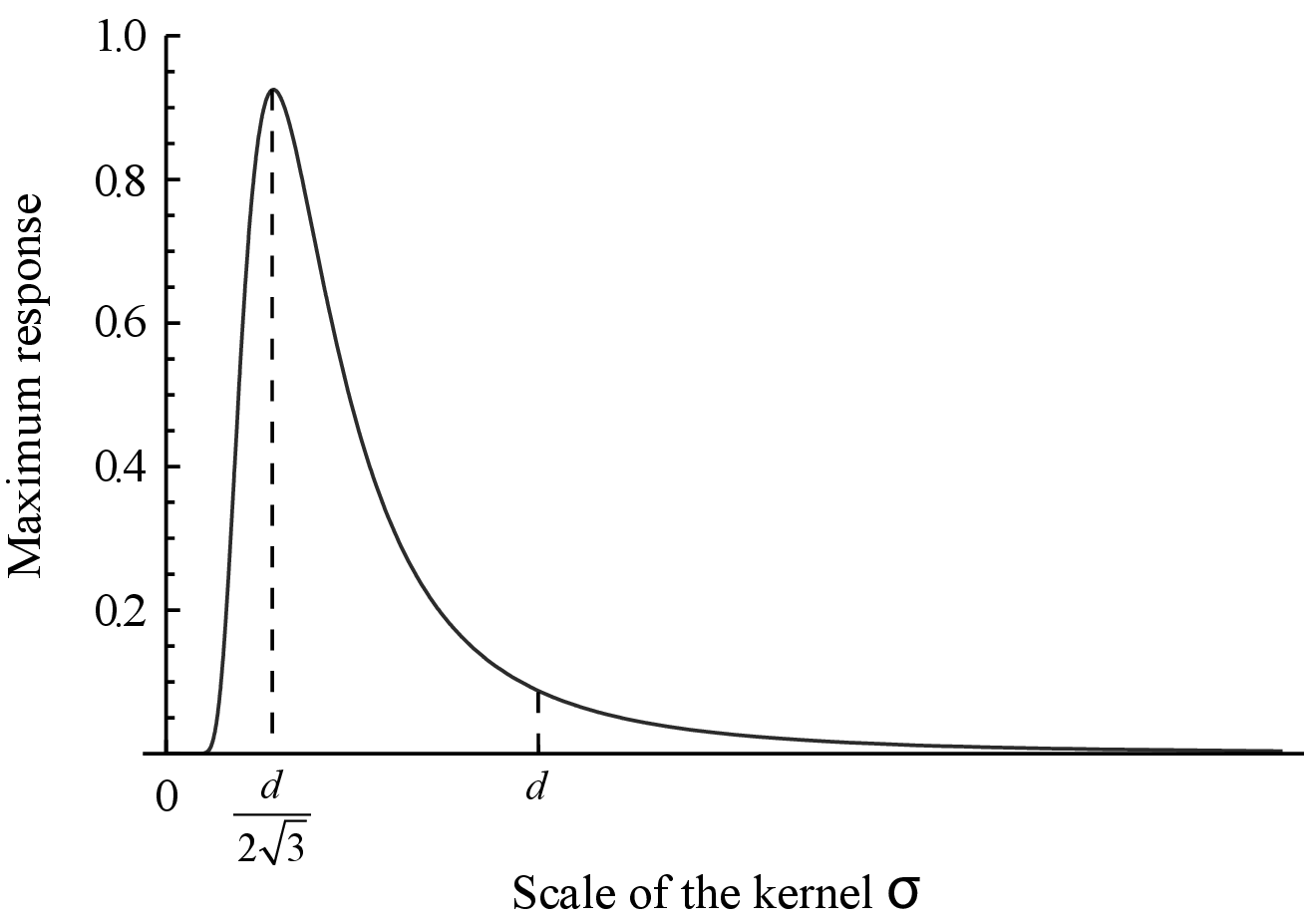}
		\caption{\label{fig:domain} Response of the normalized LoG kernels with different scales $\sigma$ to solid sphere of diameter $d$: local maximum is achieved at the value of $\sigma  = d \big/ {2\sqrt3}$.}
\end{figure}

In the three-dimensional domain the principle of multiscale filtering remains the same. The responses of the normalized LoG filter with different scales $\sigma$ to a solid sphere are shown in Figure~\ref{fig:domain}. Maximum response is reached when the scale of the kernel $\sigma$ corresponds to the diameter of the sphere $d$: $\sigma  = d \big/ 2\sqrt3$ (can be proved likewise).

\subsection {Multiscale normalized LoG filtering}

Theoretically, $\nabla_{norm}^2L(X,\sigma):\Re^3 \times \Re \to \Re$  is a real function of four continuous variables: three spatial and one scale; however, for the implementation it is also necessary to specify an appropriate  parameter quantization scheme. In our research, the quantization of spatial dimensions was selected to be identical to the quantization of the original image, i.e. the accuracy of the candidate centroid localization in each direction was limited to the voxel spacing interval along corresponding axes. 

As for the scale quantization, it is effectively the computation of $\nabla_{norm}^2G(X,\sigma)*I(X)$ for different values of $\sigma$, or "multiscale filtering." The normalized LoG response may thus be expressed as the function of discrete variables:
\begin{equation}
\nabla_{norm}^2L(X,\sigma) : S_X^3 \times S_\sigma \to \Re,
\end{equation}
where $S_X^3 = \{ 0,1,...,x_{\max}\} \times \{ 0,1,...,y_{\max }\} \times \{ 0,1,...,z_{\max }\}$ is the discrete set of three-dimensional image coordinates and $S_\sigma = \left\{\sigma_i \right\}$ is the discrete set of scales. By adding the $Lungs$ spatial constraint, the search space for nodule candidates is obtained as $(S_X^3 \cap Lungs) \times S_\sigma$.

As multiple convolutions need to be computed, the scale quantization strategy is of high importance and will affect both multiscale filter response and estimated nodule sizes. In addition, the processing of too many scales may result in unnecessary computational burden without any benefits for detection. 

In the following section we determine the design bounds for a multiscale normalized LoG filter for detecting solid spheres in a noise free situation in which both soft tissue and lung parenchyma have constant intensity values. This will set a lower bound on the parameters that will be needed for a real application in which there is image noise and some variation in tissue intensities.

\subsection{Scale quantization}
\subsubsection {a bound on maximum reduction in filter response}
\begin{figure}
		\centering
		\includegraphics[width=0.75\linewidth]{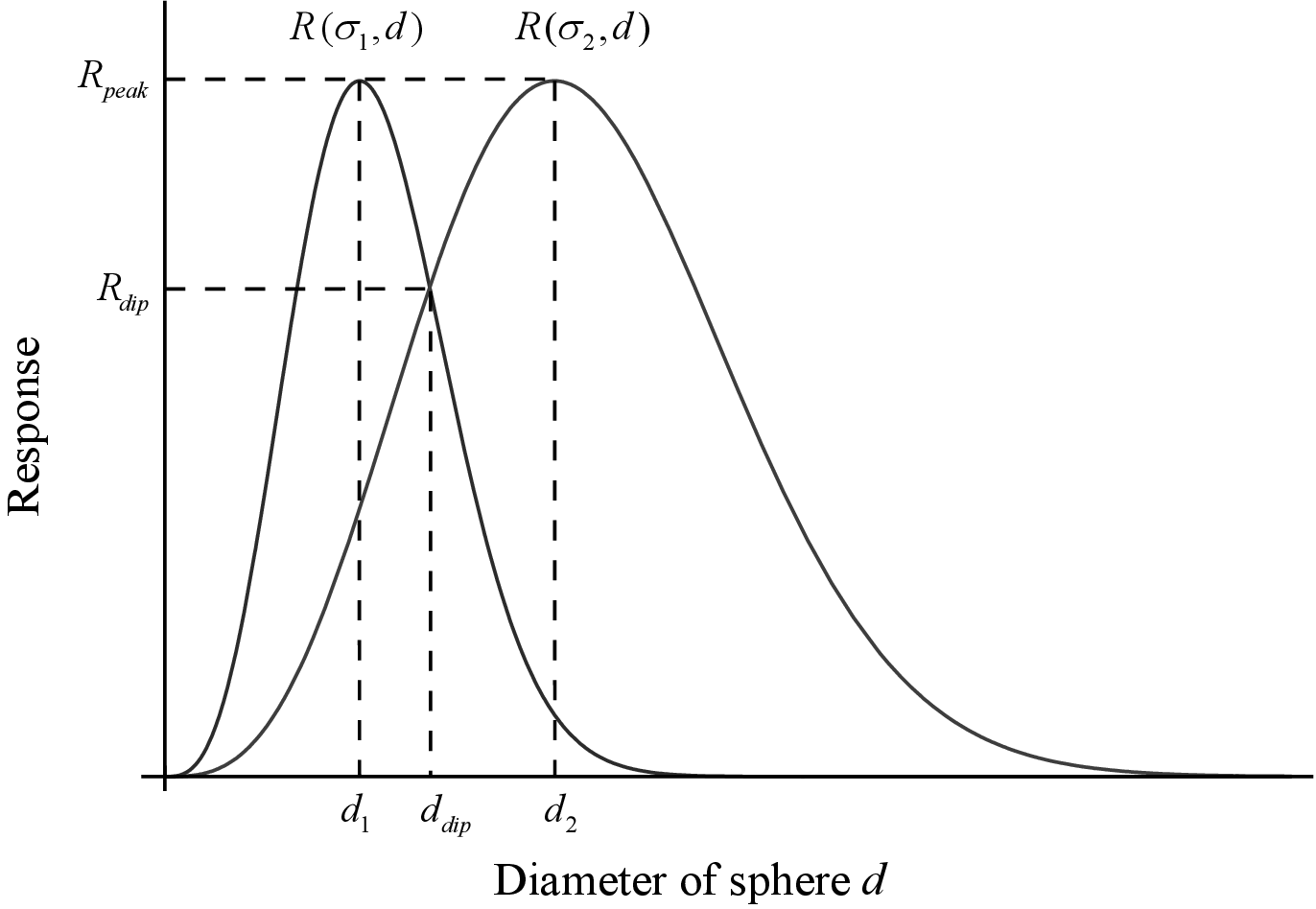}
		\caption{\label{fig:twopeaks} Response of the normalized LoG kernels of scales $\sigma_1$ and $\sigma_2$ to the solid sphere of unit intensity. Spheres with diameters $d_1 = 2 \sqrt 3 \sigma_1$ and $d_2 = 2 \sqrt 3 \sigma_2$ result in peak responses; differently sized spheres result in lower response.}
\end{figure}

The maximum response of a scale-normalized LoG filter of scale $\sigma$ to the solid sphere $S_d$ of diameter $d$ and unit intensity is reached at its center and can be found as
\begin{equation}
\label{eqn:response}
\begin{split}
R(\sigma,d)&= \int\limits_{X \in {S_d}} \nabla_{norm}^2G(X,\sigma)\;dX  = \\ 
&= \frac{d^3}{2^{2.5}\pi ^{0.5}\sigma ^3} \exp \left( \frac{-d^2}{8\sigma^2} \right).
\end{split}
\end{equation}
It can be shown, that the magnitude of the maximal response is independent of the sphere diameter and equal to
\begin{equation}
R_{peak} = \sqrt \frac{54}{\pi} \exp (-1.5) \approx 0.925.
\end{equation}
Let us consider two adjacent scales $\sigma_1$ and $\sigma_2$  ($\sigma_1 > \sigma_1$) of the multiscale filter. The sphere diameters corresponding to maximum response will be $d_1 = 2 \sqrt 3 \sigma_1$ and $d_2 = 2 \sqrt 3 \sigma_2$, respectively. These spheres will result in peak responses of the multiscale filter:
\begin{equation}
R(\sigma_1,d_1) = R(\sigma_2,d_2) = R_{peak},
\end{equation}
as shown in Figure~\ref{fig:twopeaks}.

\begin{figure}
		\centering
		\includegraphics[width=0.75\linewidth]{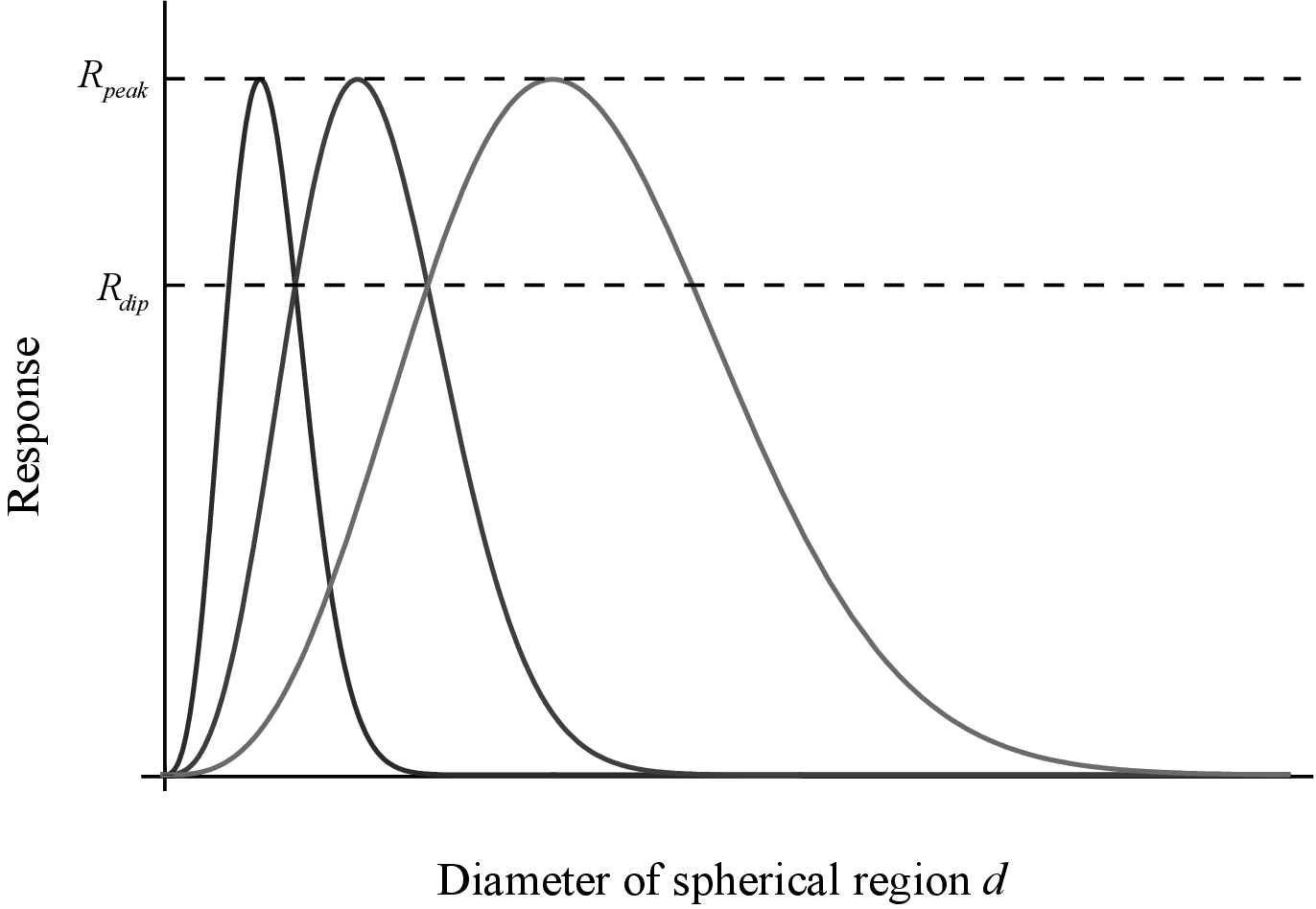}
		\caption{\label{fig:threepeaks} Exponentially increasing scale $\sigma_{i + 1} = k\sigma_i$  results in the reduction in response bounded from below by $R_{dip}$.}
\end{figure}

As the scale space is quantized, the value of maximum response will never reach $R_{peak}$, unless the diameter perfectly corresponds to a scale in the quantized set. It will result in spheres of different sizes having filter responses that diverge from the peak value.  Any sphere of diameter $d$, such that $d_1 < d < d_2$, would result in a smaller filter response. Let us estimate the diameter $d_{dip}$ that results in a minimal response $R_{dip}$ of the filter to a spherical model. Clearly, the minimum value will be reached when the responses from both scales are the same:
\begin{equation}
R_{dip} = L(\sigma_1,d_{dip}) = L(\sigma_2,d_{dip}).
\end{equation}
The nontrivial solution to this equation with respect to $d_{dip}$ is:
\begin{equation}
d_{dip} = \sqrt 8 \sigma_1\sigma_2\sqrt \frac{\ln {\sigma_2}^3 - \ln {\sigma_1}^3}{{\sigma_2}^2 - {\sigma_1}^2},
\end{equation}
which results in:
\begin{equation}
R_{dip} = \frac{4}{\sqrt \pi }\left(\sigma_1/\sigma_2 \right)^\frac{3}{1 - \left(\sigma_1/\sigma_2 \right)^2}\left( \frac{3 \ln \left(\sigma_1/\sigma _2\right)}{{\left( \sigma_1/\sigma _2 \right)^2} - 1} \right)^{1.5}.
\end{equation}
The minimal response depends on only the ratio of the adjacent scales. This means that quantization with the scale increasing in a geometric progression will result in fixed bounded error of the filter response. Therefore, to maintain a constant error bound over the range of sphere sizes of interest, we establish a set of scales with a geometric progression with a step size of $k$ with respect to the smaller scale:
\begin{equation}
\sigma_{i + 1} = k\sigma_i.
\end{equation}
For such a scale set the scale is increased exponentially, which results in an $R_{dip}$ that is independent from the scale as illustrated in Figure~\ref{fig:threepeaks}.

\subsubsection {a bound on filter response for shape confusion}
One of the criteria for selecting an appropriate scale quantization is the ability of the filter to discriminate between basic geometrical shapes. Given that pulmonary vessels are the other common structure within the lungs, we consider the behavior of a normalized LoG filter on a such structure, which is modelled as a sufficiently long solid cylinder.

The maximal response of the normalized LoG filter to a solid cylinder $C_d$ of diameter $d$ is given by
\begin{equation}
\begin{split}
R'(\sigma,d)&= \int\limits_{X \in C_d} \nabla_{norm}^2G(X,\sigma)\;dX \\
&= \frac{d^2}{4s^2} \exp\left( \frac{-d^2}{8\sigma^2} \right).
\end{split}
\end{equation}
This response reaches its maximal value of $R'_{peak}= 2 / e \approx 0.736$, when $\sigma = d \big/ 2\sqrt2$.

\begin{figure}
		\centering
		\includegraphics[width=0.75\linewidth]{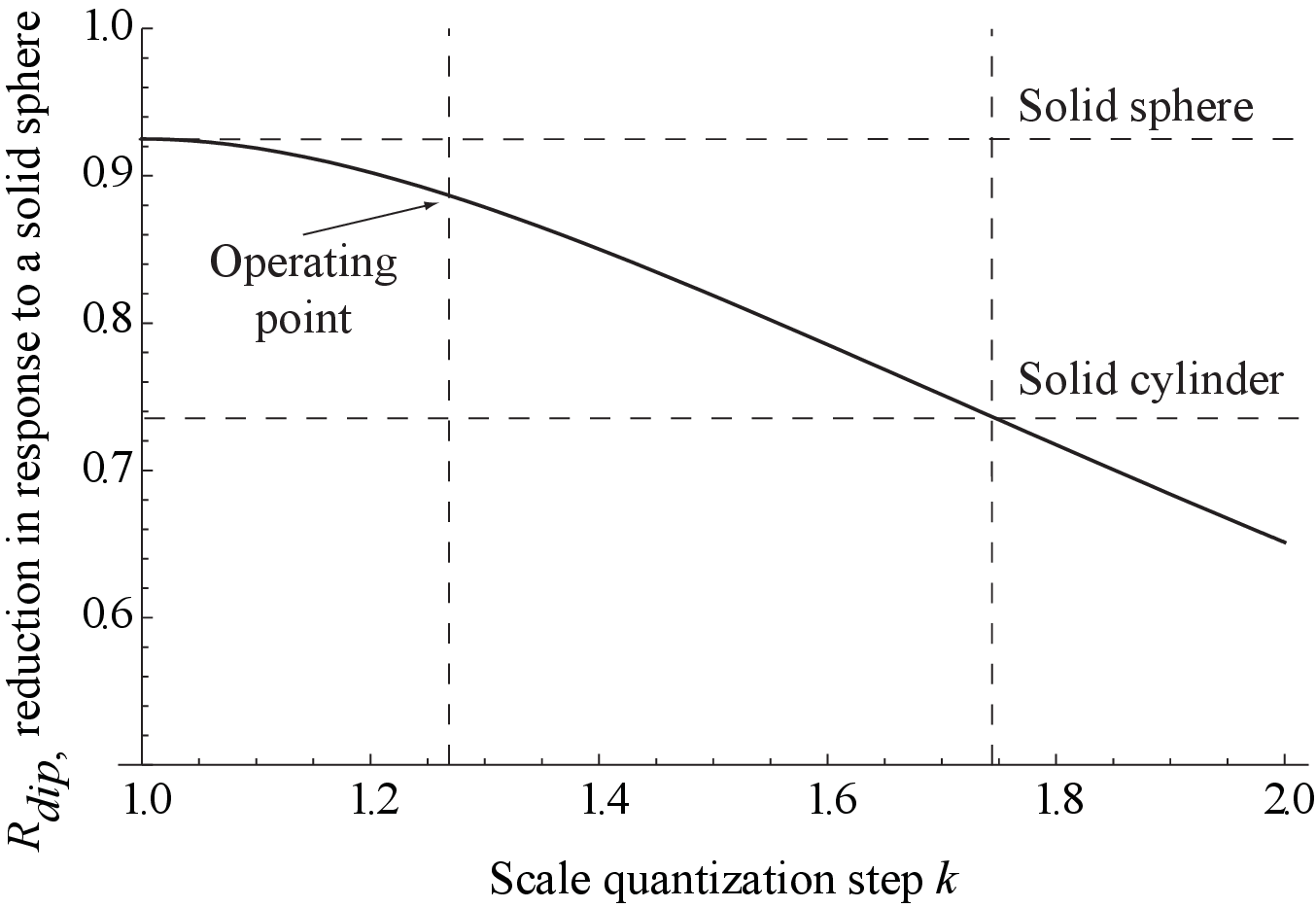}
		\caption{\label{fig:descent} Selection of operating quantization step. Solid curve shows the reduction of filter response $R_{dip}$ of the normalized LoG filter response to the solid sphere with increased quantization step $k$. Dashed lines shows the maximum responses to a solid sphere and a cylinder.}
\end{figure}

The graph in Figure~\ref{fig:descent} shows how the selection of the coefficient $k$ affects the reduction in response due to scale quantization: as the value of $k$ approaches 1.0, the value of $R_{dip}$ approaches $R_{peak} \approx 0.925$ (shown as top dashed line). This is the maximum response that can be achieved by convolving a normalized LoG kernel with a solid sphere. The bottom dashed line corresponds to the maximum response of $R'_{peak} \approx 0.736$ obtained for a solid cylinder. As the graph shows, if the scale quantization is too rough ($k > 1.746$), the multiscale filter will not be capable of reliable discrimination between solid sphere and a solid cylinder. 

\subsubsection {a bound on size estimation error}
Scale quantization has an important impact on candidate size measurement accuracy. To estimate this impact, let us consider the worst case scenario, when a solid sphere has a diameter that approaches the value of $d_{dip}$ corresponding to the maximum reduction in filter response.

\begin{figure}
		\centering
		\includegraphics[width=0.75\linewidth]{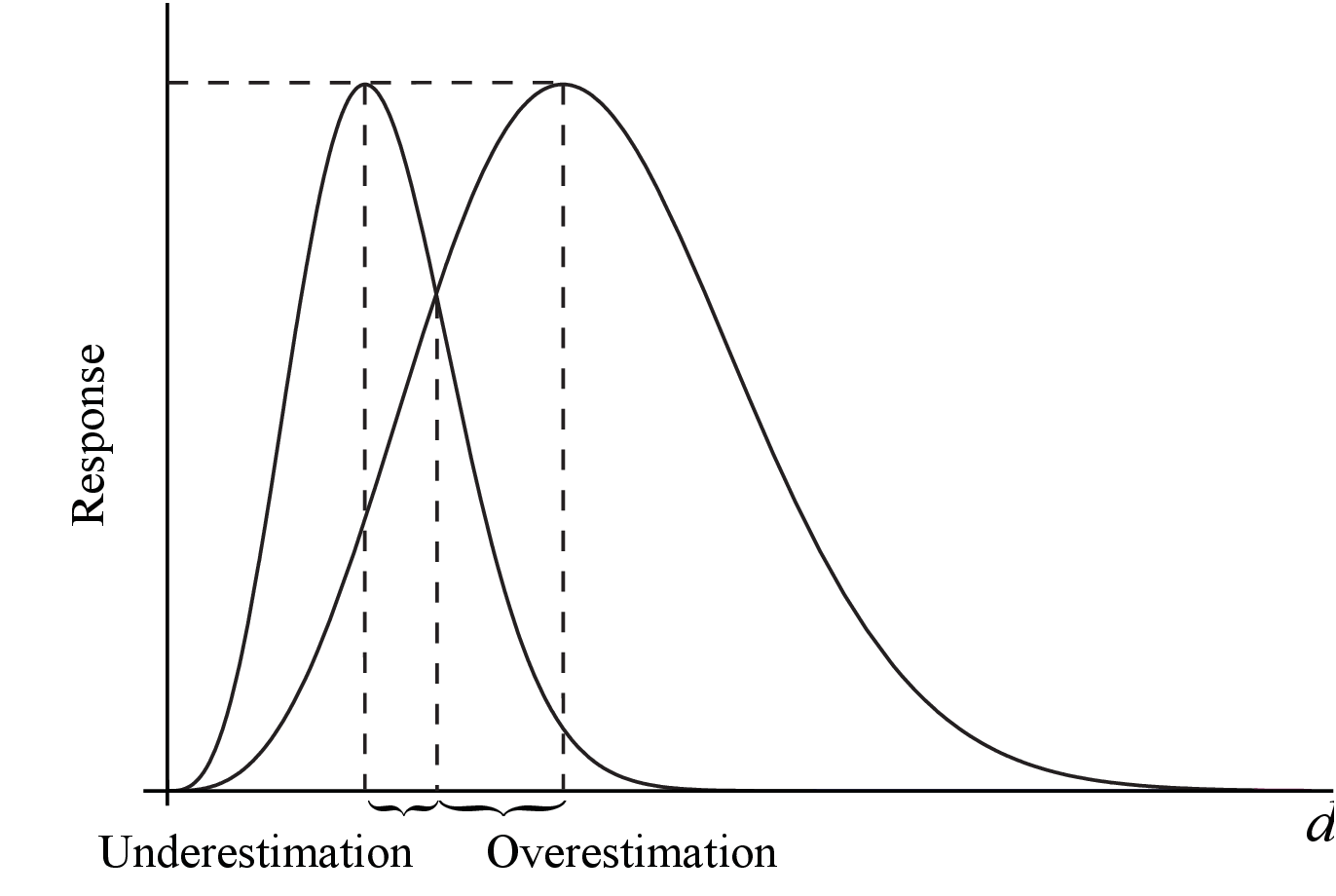}
		\caption{\label{fig:underover} Underestimation and overestimation of the sphere size due to scale quantization.}
\end{figure}

\begin{figure}
		\centering
		\includegraphics[width=0.75\linewidth]{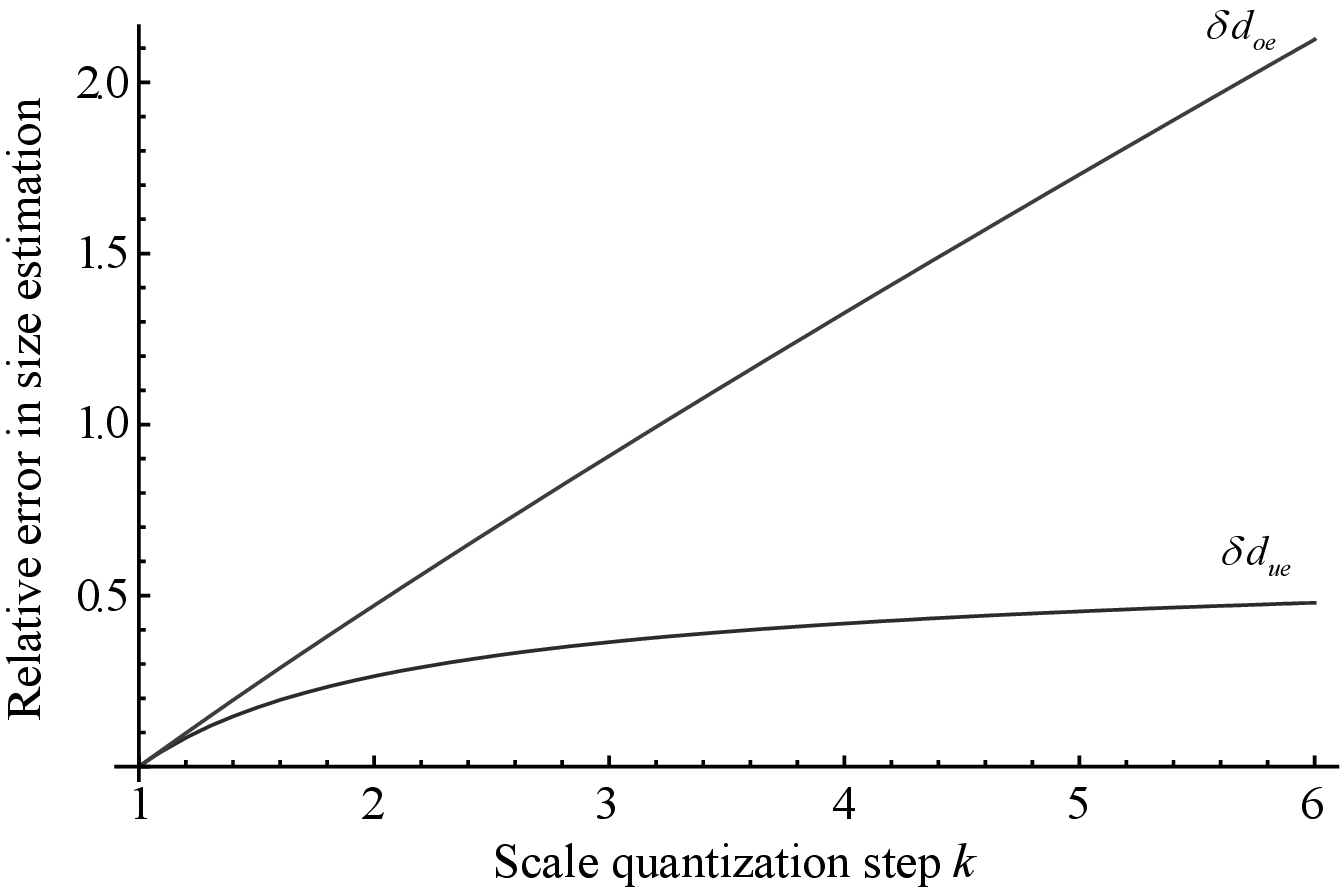}
		\caption{\label{fig:error} Relative error in solid sphere size overestimation (OE) and underestimation (UE) with respect to the size quantization step $k$.}
\end{figure}

The largest diameter underestimation (UE) will be reached when the diameter reaches $d_{dip}$ from the left side as shown in Figure~\ref{fig:underover}. 
The relative size measurement error will be equal to:
\begin{equation}
\delta d_{ue} = \frac{d_{dip} - d_1}{d_{dip}} = 1 - \sqrt {\frac{1 - k^{-2}}{2\ln k}}.
\end{equation}
The largest overestimation (OE) error, is when the diameter approaches $d_{dip}$ from the right side:
\begin{equation}
\delta d_{oe} = \frac{d_2 - d_{dip}}{d_{dip}} = \sqrt {\frac{k^2 - 1}{2\ln k}}-1.
\end{equation}
Due to the asymmetry of response function~(\ref{eqn:response}), the overestimation error is always larger than the underestimation error as shown in Figure~\ref{fig:error}.
On the one hand, the values of $k$ close to 1 result in both less filter confusion and better size estimation. On the other hand, with such a small step size, the number of scales needed to cover the nodule size range of interest will be high, which is not desirable from the computational cost point of view.

In our experiments, the value of $k = 1.27$ and 10 scales corresponding to sphere diameters increasing in geometrical progression from 3 to 25 mm were used. This was done to match the size range of targeted nodules plus two boundary scales. The main reasons for selection of this quantization strategy were maintaining a bound on reduction in LoG filter response and a bound on shape confusion. In this configuration, the highest reduction in response results in $R_{dip} \approx 0.887$. Relative errors in size underestimation and size overestimation were no greater than $\delta d_{ue} = 0.11$ and $\delta d_{oe} = 0.13$, respectively. The set of quantized diameters $d_i$, corresponding scales $\sigma_i$, and effective diameter ranges are shown in Table~\ref{tab:quantization}. All spheres in each $i-th$ effective diameter range (fourth column) will be detected as having diameter $d_i$.
 
\begin{figure}
		\centering
		\includegraphics[width=0.75\linewidth]{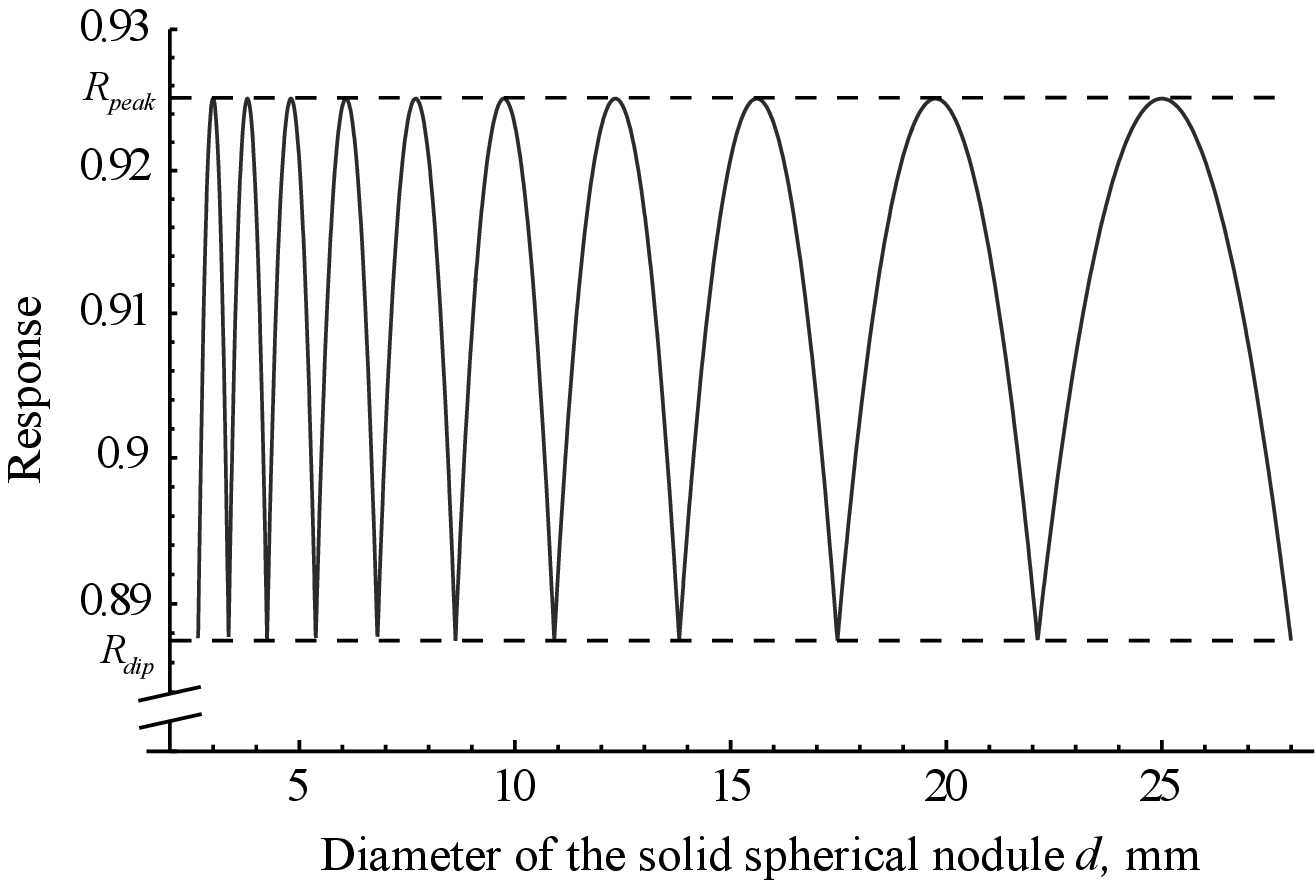}
		\caption{\label{fig:multiresponse} Responses of the multiscale LoG filter to three-dimensional solid spheres of different diameters.}
\end{figure}

\begin{table}
\renewcommand{\arraystretch}{1.3}
\caption{Quantization of candidates sizes. The columns are: scale index, diameter of the kernel, corresponding scale and range of the candidate sizes assigned to the scale.} 
\label{tab:quantization}
\begin{center}
\small
\begin{tabular}{|l|l|l|l|}
\hline
$i$ & $d_i$, mm & $\sigma_i$ & $range_i$, mm \\
\hline
0 (boundary scale) &	2.37&	0.68&	N/A\\
1	&3.00	&0.86	&2.65  -  3.35\\
2	&3.79	&1.09	&3.35  -  4.25\\
3	&4.80	&1.38	&4.25  -  5.38\\
4	&6.08	&1.75	&5.38  -  6.81\\
5	&7.69	&2.22	&6.81  -  8.62\\
6	&9.74	&2.81	&8.62  -  10.91\\
7	&12.33	&3.55	&10.91  -  13.80\\
8	&15.60	&4.50	&13.80  -  17.47\\
9	&19.75	&5.70	&17.47  -  22.11\\
10	&25.00	&7.21	&22.11  -  27.99\\
11 (boundary scale)	&31.64	&9.13	& N/A\\
\hline
\end{tabular}
\end{center}
\end{table}
Corresponding response function of the multiscale LoG filter is illustrated in Figure~\ref{fig:multiresponse}.

\subsection {The impact of spatial interference}

In previous sections, the scale quantization effects and properties of our nodule model in isolation from other pulmonary structures were reviewed. For a better understanding of the behavior of the LoG filtering in practical sense, it is important to estimate the stablity of the nodule model in the presence of other spatial structures such as solid cylinder and wall that closely approximate a pulmonary vessel and chest wall, respectively.

\begin{figure}
		\centering
		\includegraphics[scale=0.5]{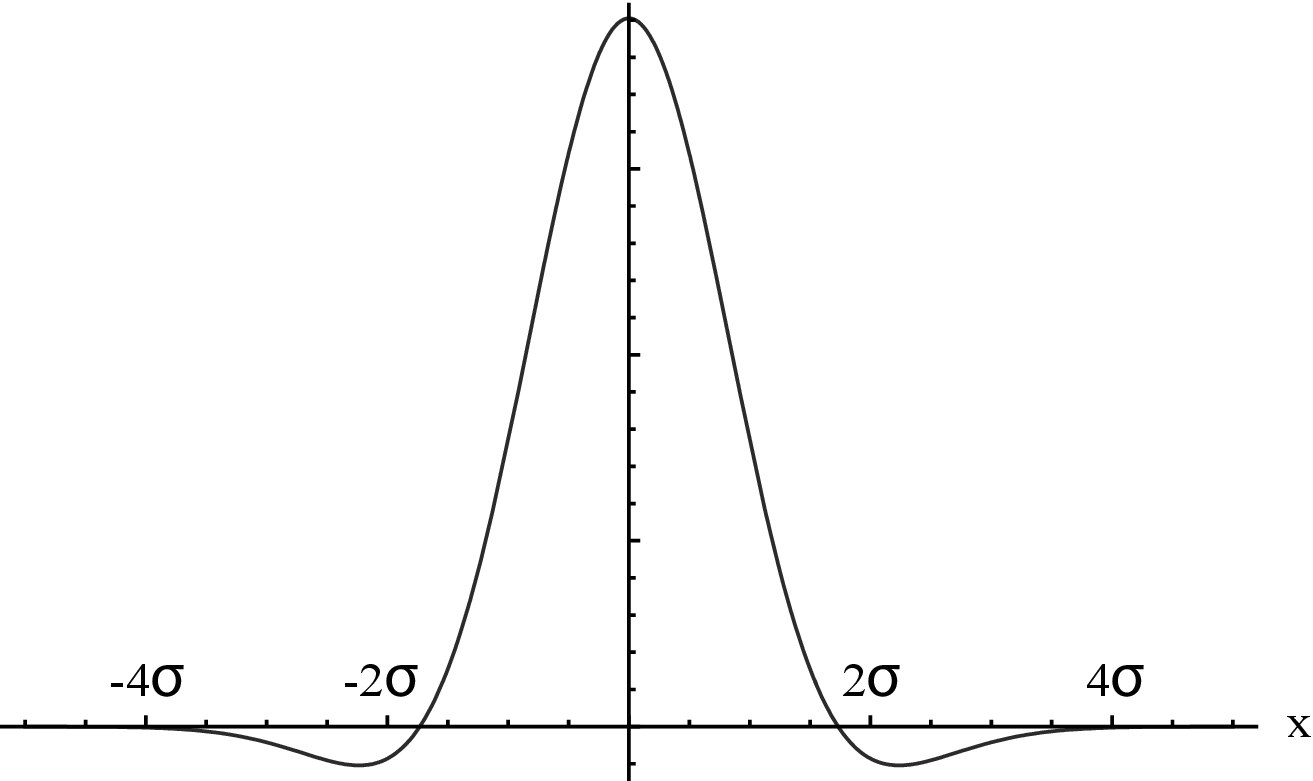}
		\caption{\label{fig:log} Central slice of the three-dimensional normalized LoG kernel centered at the origin of coordinate system. Major part of the volume of support is enclosed within the radius of $4\sigma$ from the origin.}
\end{figure}

For example, if a nodule is located close to a pulmonary vessel or "attached" to a chest wall, the response of the filter to the nodule would be altered due to superposition in the response to the interfering structure. This effect is amplified as the objects get closer to each other. From the profile of the negated LoG kernel centered at the origin of the coordinate system, shown in Figure~\ref{fig:log}, one may infer that all significant nonzero values are concentrated near the origin and do not extend beyond $4\sigma$ (or slightly greater than doubled zero-crossing) distance. This means that the objects located closer than $4\sigma$ from each other may produce a noticeable interference effect.

To quantify the interference effect, two simulations for an observed solid sphere model of unit diameter and intensity were conducted. The purpose of these simulations was to imitate the most common interference happening within the lungs: with a pulmonary vessel and with a chest wall. In the first simulation we recorded how a proximal long solid cylinder of the same diameter and intensity affects the sphere filter response and size estimate. In the second simulation the solid sphere was placed near a solid flat wall of sufficiently large size and thickness. The simulations were conducted using discrete synthetic three-dimensional images of these objects. For each distance (measured in sphere diameters), the normalized LoG space maxima corresponding to the sphere was found and the best scale (size estimate) and filter response were recorded.

\begin{figure}
		\centering
		\includegraphics[scale=0.4]{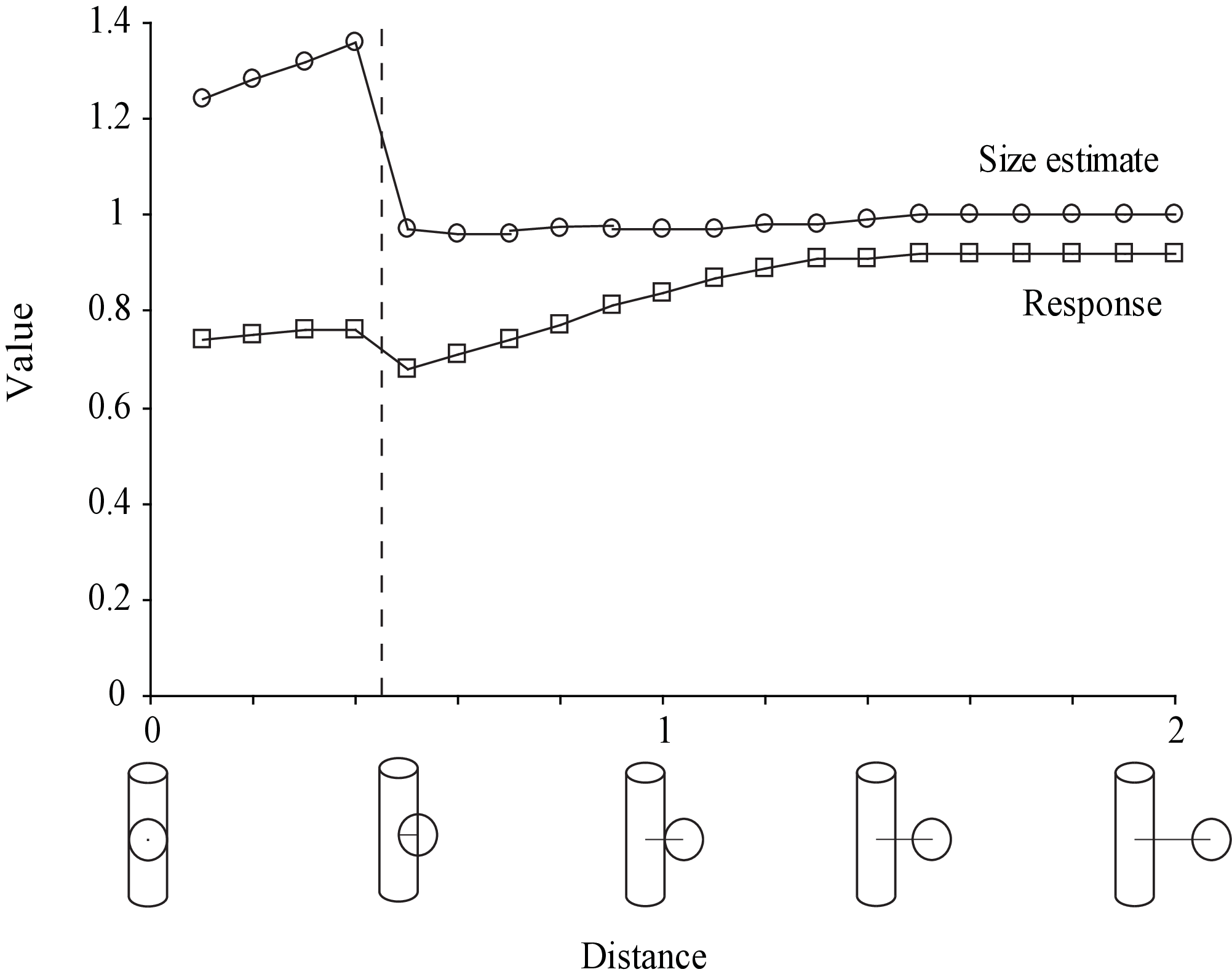}
		\caption{\label{fig:cylsphere} Cylinder-sphere interference. Effect of the distance between the objects (measured in sphere diameters) on filter response and sphere size estimate. Dashed line indicates the border between detecting one single "blob" and two separate "blobs."}
\end{figure}

\begin{figure}
		\centering
		\includegraphics[scale=0.4]{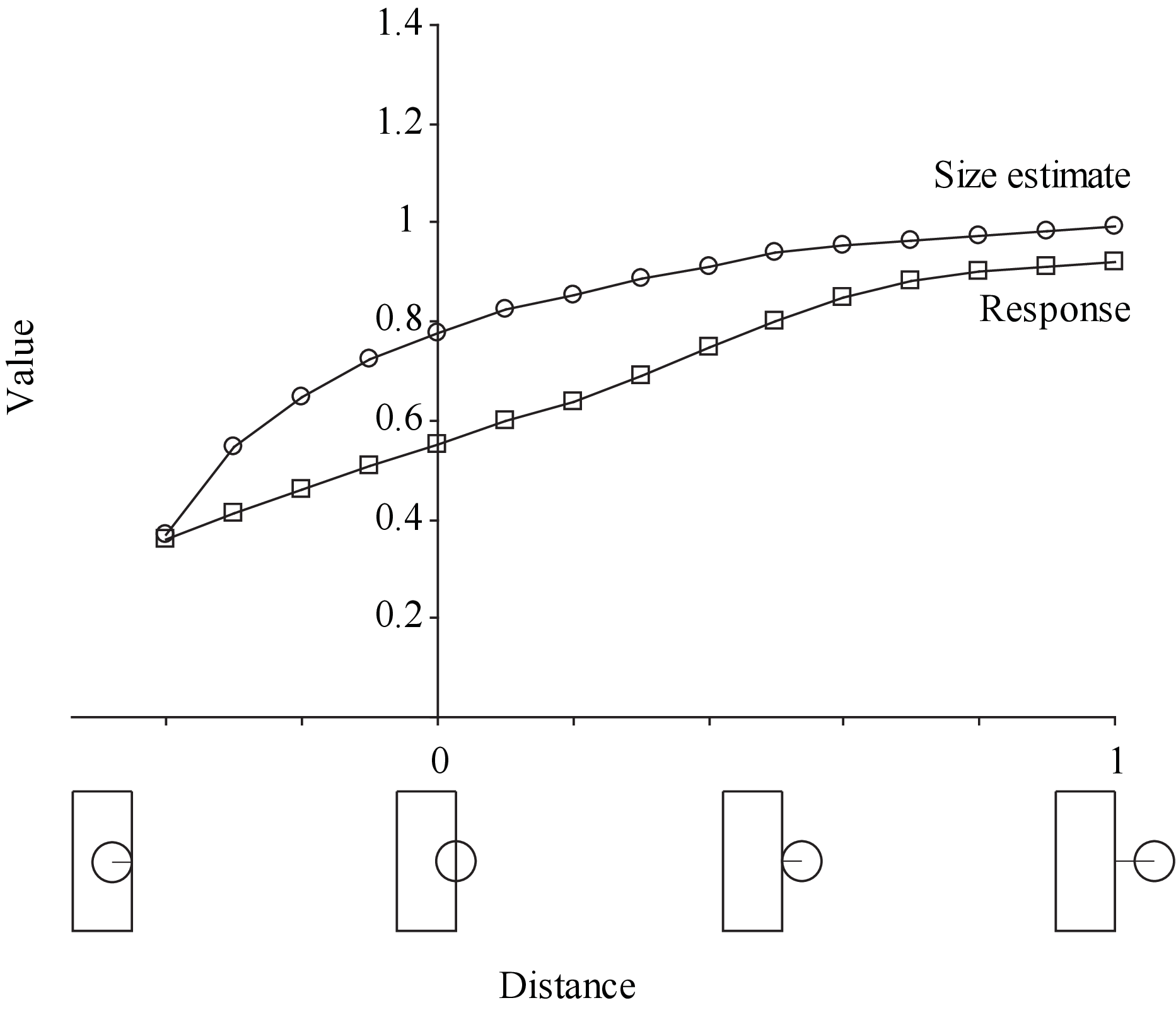}
		\caption{\label{fig:sphwall} Sphere-wall interference. Effect of the distance between the centroid of a sphere and the wall (measured in sphere diameters) on the filter response and size estimate.}
\end{figure}

Results of the sphere-cylinder interference simulation are shown in Figure~\ref{fig:cylsphere}. When the distance from the central axis of the cylinder to the center of the sphere was greater than 1.5, no negative effects from interference were observed. As the distance between the objects is reduced, the response of the filter decreases. The size estimate maintained close to 1.0 until the objects started to form an overlap. Additional decreasing of the distance to 0.5 resulted in a decrease of response. Finally, after passing the 0.5 threshold, the normalized LoG filter was not able to distinguish between the sphere and the cylinder and detected a single "blob" instead; a jump in both response and size estimate was observed. Finally, when the objects overlay entirely, the filter "converges" to the response and size estimates of an ideal standalone cylinder.
The effect of sphere-wall interference is shown in Figure~\ref{fig:sphwall}. Here, as the sphere approaches and merges with the wall, its estimated size is decreased. Even though there is no unambiguous definition of the "size" for a sphere partially attached to a wall, it is reasonable to assume that it should be smaller than the one of the isolated sphere. Similarly, as the sphere is "immersed" into the wall, the response decreases in agreement with an increased degree of attachment.

In both scenarios, the sphere, unless merged with the interfering object entirely, resulted in a distinct maximum in the normalized LoG search space. In spite of diminished response and altered size for proximity distances less than 0.5, this can be resolved by a later stage of a CAD system that may still classify this candidate as a true positive.

These simulations confirmed that the interference exists and affects the filter response and size estimate of proximal objects.  The simulations also showed that the interference effect is within reasonable limits and should not cause major obstacles to candidate identification. However, if the adjacent object has an intensity that is higher than the target nodule, the interference effect may be amplified, e.g. in the case of detecting nonsolid nodules in proximity to solid structures. This is discussed later in the paper in the nonsolid candidate generation section.

\subsection{Minimum response criterion}

The proposed multiscale LoG filtering scheme would produce local maxima in the search space for both true nodules and other image structures including noise. As a post-processing step we propose to threshold and eliminate the low-response candidates that do not meet our solid nodule model, both shape- and intensity-wise. 

To determine mimimum response threshold it is necessary to consider the "worst" scenario where a true nodule would have a minimum possible response. This may happen when the response to the nodule is degraded due to all of the following factors all together: (a) interference from other pulmonary structure; (b) quantization of detection scales; (c) low tissue contrast.

To model a situation such as this, let us refer to the cylinder-sphere interference simulation experiment, described above. The greatest reduction in response due to interference results in the response for sphere equal to: $R_{int} \approx 0.7$ as shown in Figure~\ref{fig:cylsphere}. If we consider the reduction in response due to quantization, the "worst" response will be further reduced by a factor of: $R_{dip} / R_{peak} = 0.958$. Here we made an assumption that the reduction of response due to quantization in the case of interference is of the same order as the one computed for an isolated sphere. 

Previously we considered the objects having intensity of 1.0 and backround having intensity of 0.0. Using the linearity of the LoG filtering, one can recompute the response function in the real CT scan domain. The median intensity for lung parenchyma tissue is $I_{paren}$ = \mbox{-810 HU}~\cite{WBWB07} and the lower bound on solid tissue intensity is $I_{solid,\min}$ = \mbox{-474 HU}~\cite{browder2007automated}. Therefore, the final "worst" response can be written as
\begin{equation}
R_{CT} = R_{int}\frac{R_{dip}}{R_{peak}} \cdot \left(I_{solid,\min } - I_{paren}\right) \approx 226.
\end{equation}
In other words, the scenario where solid nodule of lowest contrast is affected by both maximal spatial interference and the extreme scale quantization would result in a LoG filter response of 226. The values of responses greater than this threshold value are sufficient for detecting solid nodules and filtering out noise and structures other than nodules.

One may hypothetically think of an even worse situation where a sphere is merged with a solid wall of greater intensity. This situation may happen if there is a rib that is close to a large nodule "attached" to the chest wall. In this case, the response threshold must be lowered, or, alternatively, the high intensity rib may be suppressed from the image by windowing; similar technique is considered in subsection~\ref{nonsolid_section}.

\subsection{Candidate generation scheme}

The main steps for the multiscale normalized LoG solid nodule candidate generator are the following:
\begin{enumerate}
	\item Segment the spatial mask $Lungs$ from the CT image.
	\item Compute $\nabla_{norm}^2L(X,\sigma_i)$ response function for discrete set of scales.
	\item Identify nodule candidates from given $Lungs$ and $\nabla_{norm}^2L(X,\sigma)$.
	\item Delete candidates with low filter response.
\end{enumerate}

The first step of the algorithm involves computation of the lungs spatial mask that will limit the spatial search space of the algorithm. It was obtained as described in the search space demarcation section of Enquobahrie et al.~\cite{AEAR07}. Prior to candidate generation, the lungs region was extended outwards by morphological dilation with a solid sphere of 10 mm. The purpose of such an extension was to account for lung segmentation imperfections and to make sure that the segmented lungs will encompass all nodules.

The main purpose of the second step is to compute the response function $\nabla_{norm}^2L(X,\sigma_i)$ for a discrete set of scales. To optimize the convolution operation, the convolution theorem was used, while the computation is carried in the frequency domain. An outline of the optimization is given in Appendix~\ref{apx1}, while the scheme of the entire step is shown in Algorithm~\ref{alg:alg1}. In short, to find the response for each of the scales, the Discrete Fourier Transforms (DFT) of the original image was found and multiplied with a pre-computed transform of the normalized LoG function; then the inverse DFT was taken and normalized. For computation of the DFTs, the FFTW library~\cite{frigo2005design} was used. It allows computation of the convolution of two 512x512x512 images within a few seconds.

\begin{algorithm*}
\def\la{\leftarrow}
\def\FOR{\textbf{for} } 
\def\FROM{\textbf{ from } } 
\def\FOREACH{\textbf{for each } }
\def\TO{\textbf{to} }
\def\STEP{\textbf{step} }
\def\DO{\textbf{do} }
\def\ENDFOR{\textbf{end for} }
\def\IF{\textbf{if} }
\def\THEN{\textbf{then} }
\def\ENDIF{\textbf{end if} }
\def\GOTO{\textbf{go to} }
\def\STOP{\textbf{stop} }
\caption{Constructing discrete response function}
\label{alg:alg1}
\medskip
\begin{tabular}{l l}
$S_\sigma$ & discrete set of scales \\
$I(X)$ & original image
\end{tabular}
\medskip\hrule\medskip
\begin{tabular}[t]{l l}
$I_\mathcal{F}(\Omega) \la \mathcal{F}\{I(X)\}$ & find the DFT of original image\\
$\FOREACH \sigma_i \in S_\sigma$	& for each discretization step \\
\quad $M(\Omega, \sigma_i) \la I_\mathcal{F}(\Omega) \cdot \mathcal{F}\left\{{\nabla^2 G(X, \sigma_i)}\right\}$ & multiply the image DFT and precomputed LoG kernel DFT\\
\quad $\nabla^2L(X,\sigma_i) \la \mathcal{F}^{-1} \left\{ M(\Omega,\sigma_i) \right\}$ & find the inverse DFT\\
\quad $\nabla_{norm}^2L(X, \sigma_i) \la -\sigma_i^2 \nabla^2L(X, \sigma_i)$ & normalize\\
\ENDFOR
\end{tabular}
\end{algorithm*}

\begin{figure*}
\centering
\subfigure[Original image]{\label{fig:i0}\includegraphics[scale=0.2]{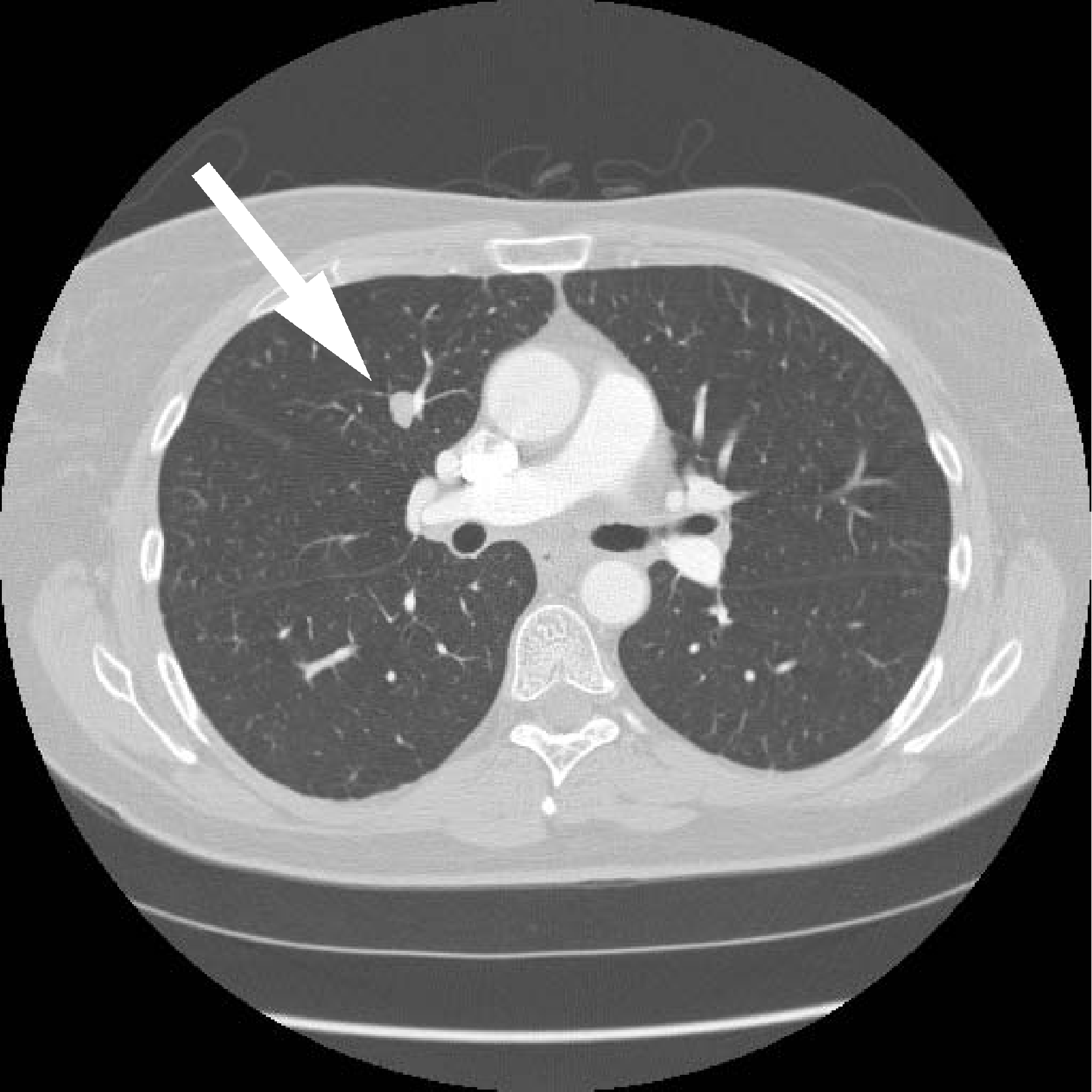}}
\subfigure[$\sigma_1=0.86$]{\label{fig:i1}\includegraphics[scale=0.5]{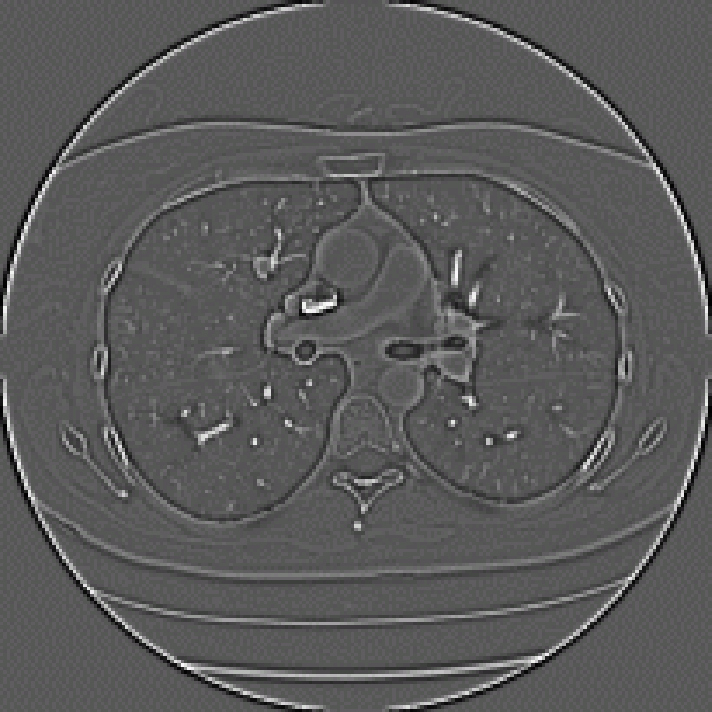}}
\subfigure[$\sigma_2=1.09$]{\label{fig:i2}\includegraphics[scale=0.5]{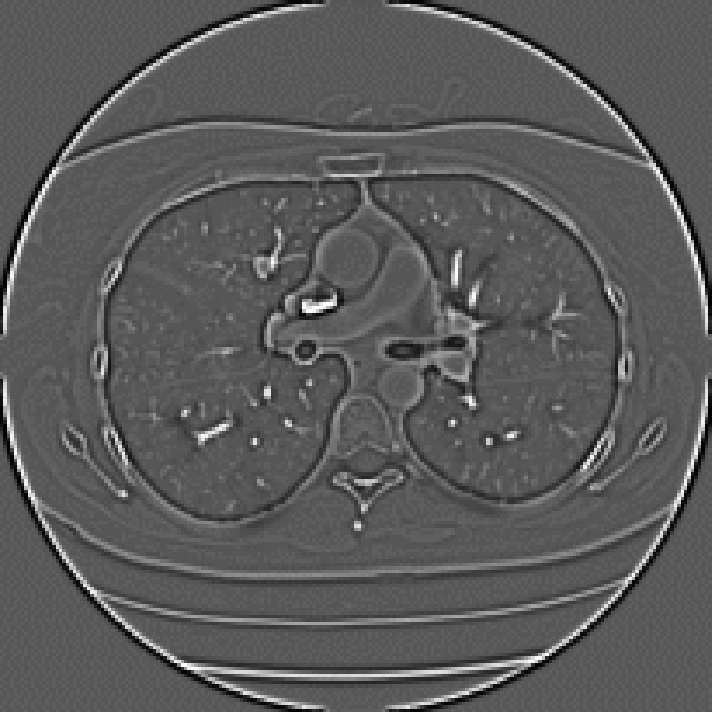}}
\subfigure[$\sigma_3=1.38$]{\label{fig:i3}\includegraphics[scale=0.5]{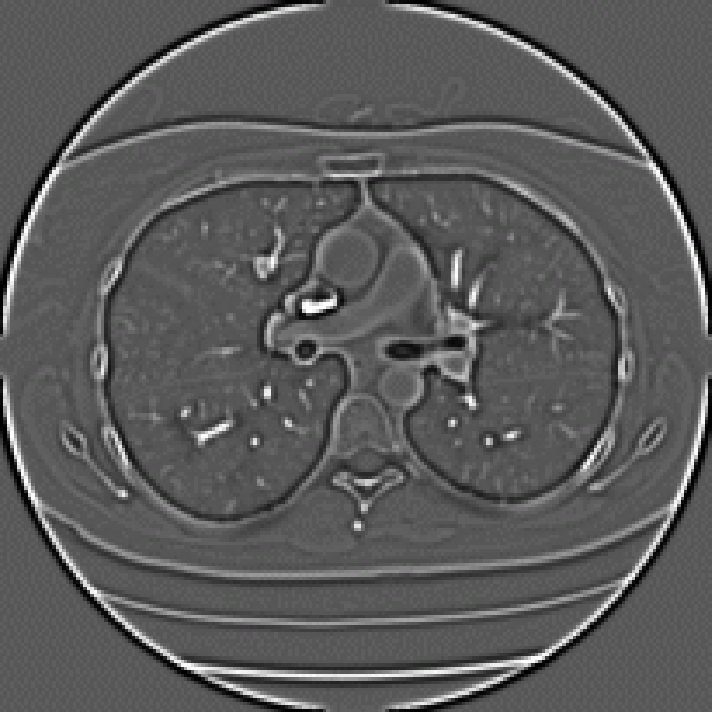}}
\subfigure[$\sigma_4=1.75$]{\label{fig:i4}\includegraphics[scale=0.5]{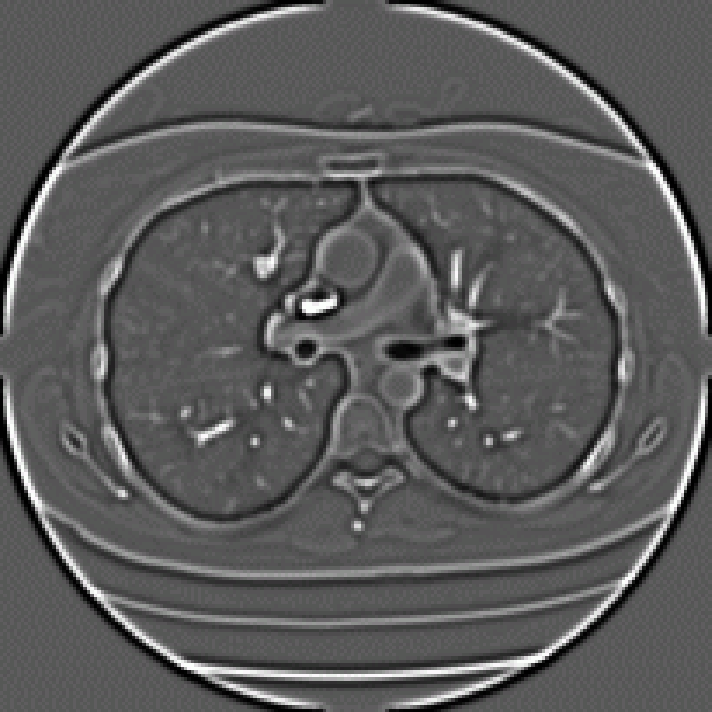}}
\subfigure[$\sigma_5=2.22$]{\label{fig:i5}\includegraphics[scale=0.5]{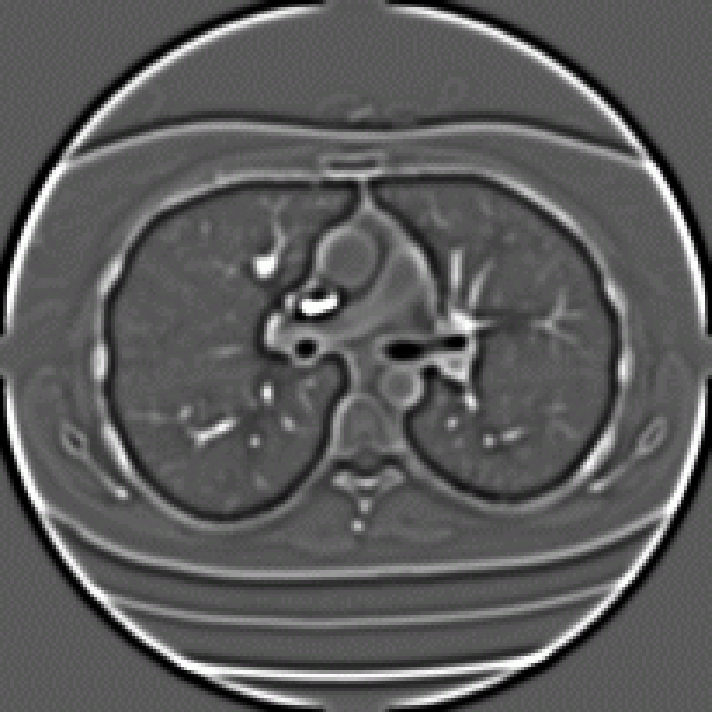}}
\subfigure[$\sigma_6=2.81$]{\label{fig:i6}\includegraphics[scale=0.5]{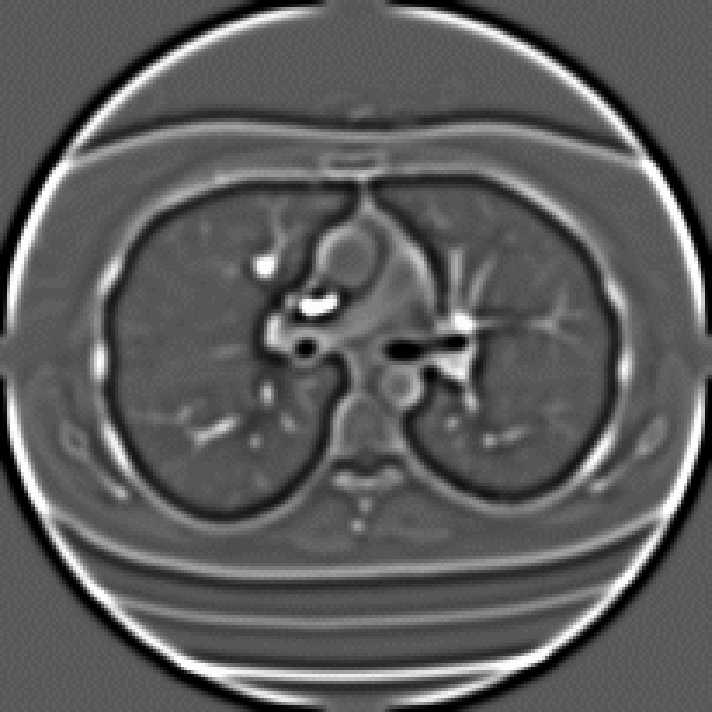}}
\subfigure[$\sigma_7=3.55$]{\label{fig:i7}\includegraphics[scale=0.5]{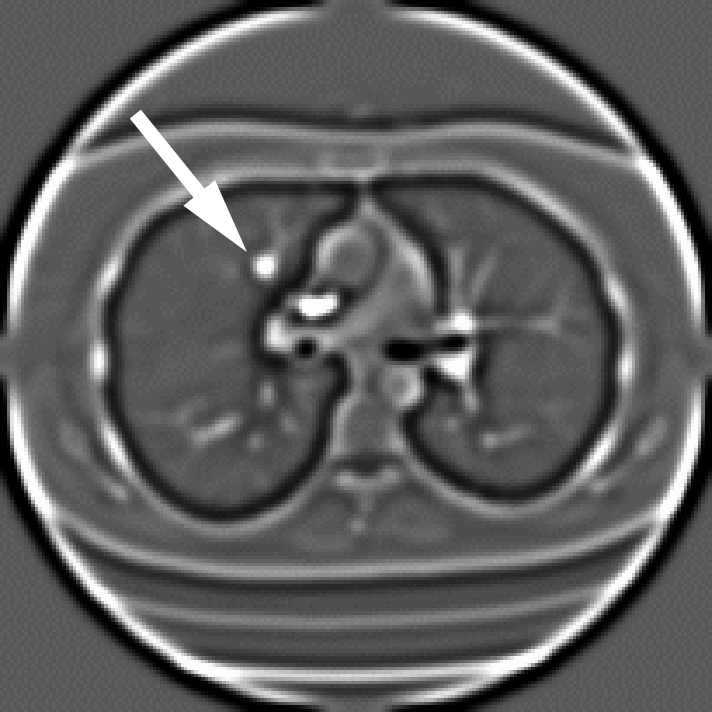}}
\subfigure[$\sigma_8=4.50$]{\label{fig:i8}\includegraphics[scale=0.5]{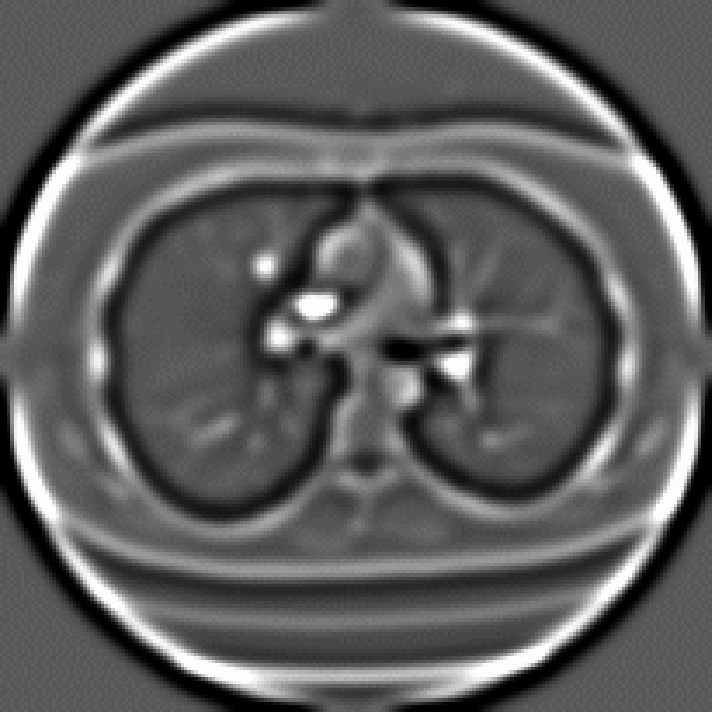}}
\subfigure[$\sigma_9=5.70$]{\label{fig:i9}\includegraphics[scale=0.5]{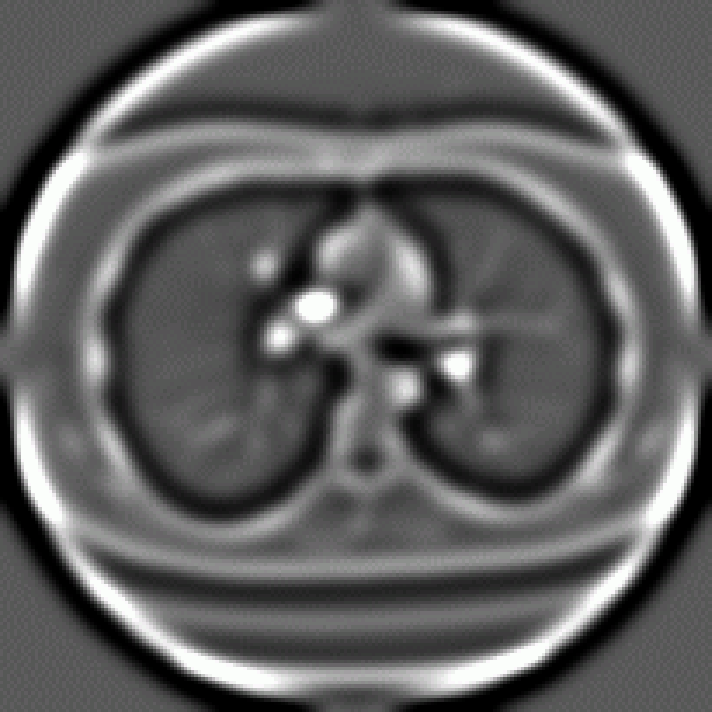}}
\subfigure[$\sigma_{10}=7.21$]{\label{fig:i10}\includegraphics[scale=0.5]{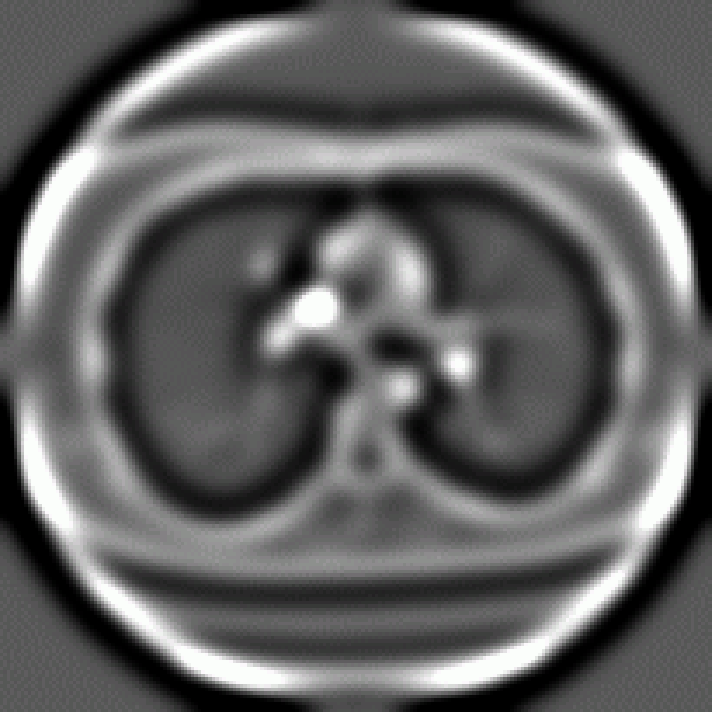}}
\caption{Response functions at different scales. Shown is the original CT scan (a) and computed responses (b) - (k). For simplicity of visualization, only one two-dimensional slice of the three-dimensional response function is shown for each of the scales. The maximum response to the nodule is achieved at the scale $\sigma_7 = 3.55$ corresponding to the size of the nodule $d = 12.33\ mm$.}
\label{fig:resp10}
\end{figure*}

The response function for the example CT image is shown in Figure~\ref{fig:resp10}. It consists of multiple response functions computed for each scale.

The third step involves finding the candidates given the set of search subspaces for each scale. Algorithm~\ref{alg:alg2} illustrates the procedure.

\begin{algorithm*}
\def\la{\leftarrow}
\def\FOR{\textbf{for} } 
\def\FOREACH{\textbf{for each} }
\def\TO{\textbf{to} }
\def\STEP{\textbf{step} }
\def\DO{\textbf{do} }
\def\ENDFOR{\textbf{end for} }
\def\IF{\textbf{if} }
\def\THEN{\textbf{then} }
\def\ENDIF{\textbf{end if} }
\def\GOTO{\textbf{go to} }
\def\STOP{\textbf{stop} }
\caption{Constructing nodule candidates set $C$}
\label{alg:alg2}
\medskip
\begin{tabular}[t]{l l}
$\nabla_{norm}^2L(X, \sigma)$ & computed discrete response function\\
$S_X^3$ & discrete image search space\\
$S_\sigma$ & discrete set of scales \\
$Lungs$ & spatial lungs mask
\end{tabular}
\medskip\hrule\medskip
\begin{tabular}[t]{l l}
$C \la \emptyset$ & start with empty set of nodule candidates \\
\FOREACH $\sigma_i \in S_\sigma$ & for each scale \\
\quad \FOREACH $(Y,\sigma_i) \in (S_X^3 \cap Lungs) \times S_\sigma$	&for each discrete point in the search space\\
\quad\quad \IF $\nabla_{norm}^2L(Y, \sigma_i) = \max \left\{ \nabla_{norm}^2L(Z, \zeta) : (Z, \zeta) \in \mathcal{N}(Y, \sigma_i) \right\}$& if the point is search space local maxima\\
\quad\quad\quad $C \la C \cup (Y, \sigma_i)$ & update the candidates set \\
\quad\quad \ENDIF & \\
\quad \ENDFOR & \\
\ENDFOR & \\
\end{tabular}
\end{algorithm*}

\begin{figure}
		\centering
		\includegraphics[width=0.95\linewidth]{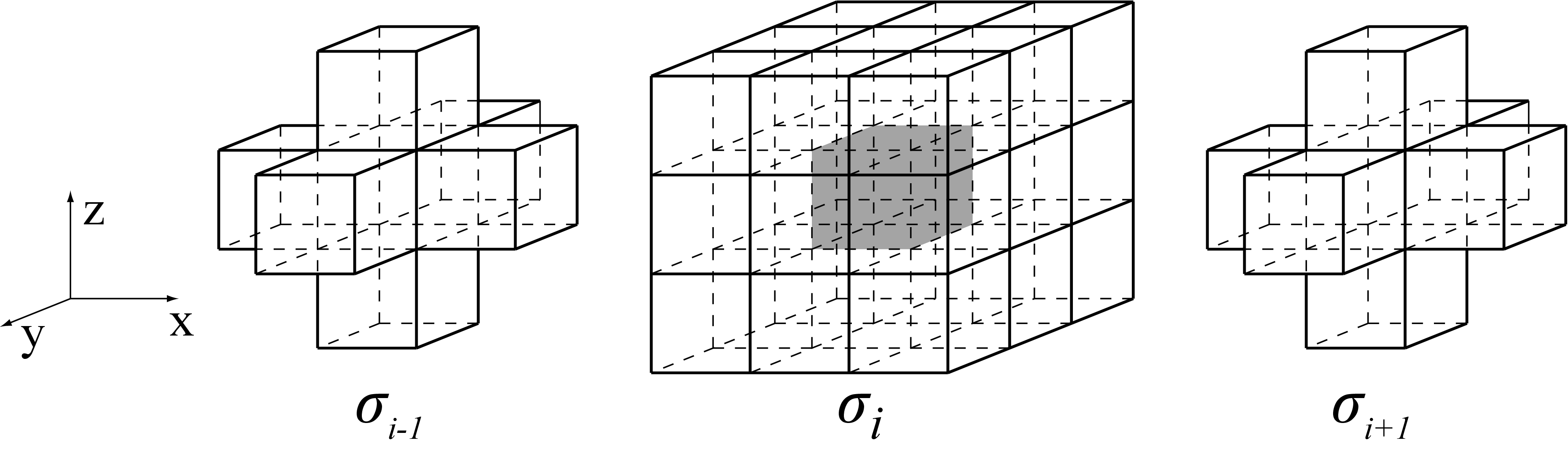}
		\caption{\label{fig:cubes} Four-dimensional neighborhood around a sample point in the discrete search space.}
\end{figure}

The third step of the algorithm finds local maxima in the four-dimensional search space (except boundary scales) with respect to both location and scale. Local maxima are determined by comparing of the filter response in a given point $(Y,\sigma_i)$ to the responses of all its neighboring points $(Z,\zeta) \in \mathcal{N}(Y,\sigma_i)$. The considered neighborhood included 26 adjacent points on the discrete grid at the current scale and 7 adjacent points at scales above and below as shown in Figure~\ref{fig:cubes}.

If the filter response at a considered point was maximal, i.e. it was greater or equal to the responses of all other neighboring points, the point was included to the set of nodule candidates. From the implementation convenience, instead of performing full four-dimensional search, the scales were processed in sequence while keeping in memory only the current scale and one scale above and below. 

\begin{figure}
\centering
\subfigure[]{\label{fig:deletionbefore}\includegraphics[width=0.95\linewidth]{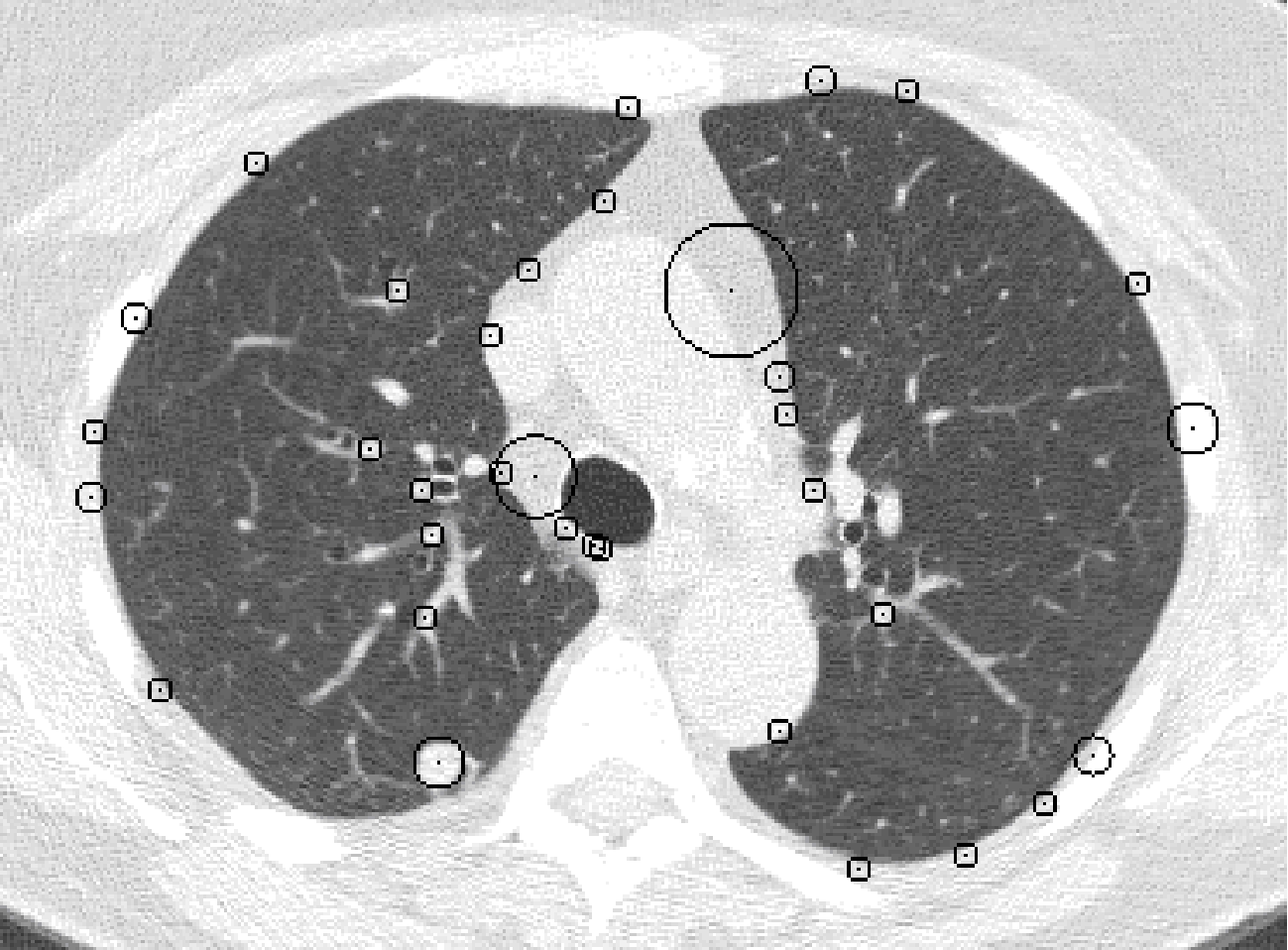}}
\subfigure[]{\label{fig:deletionafter}\includegraphics[width=0.95\linewidth]{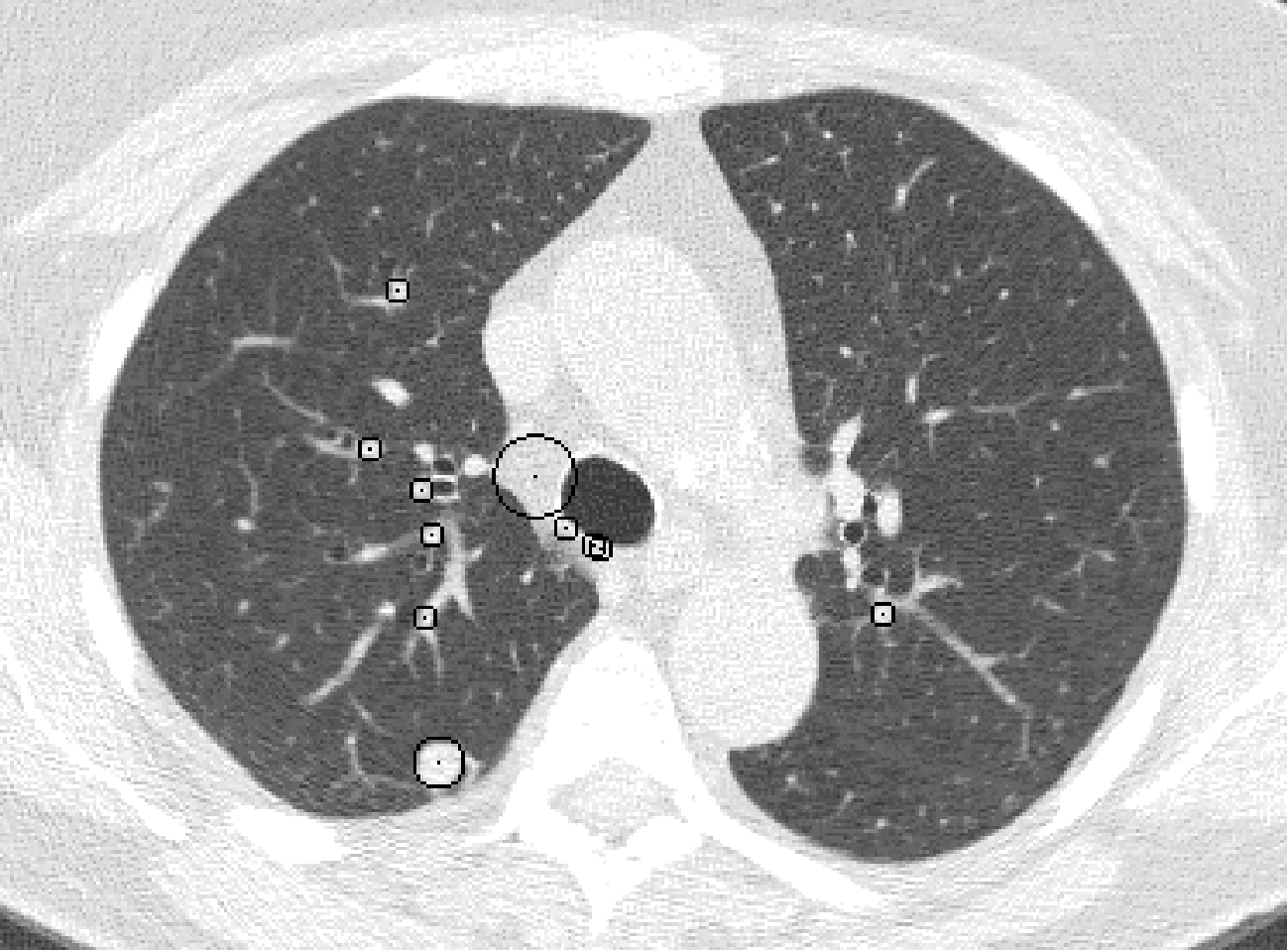}}
\caption{An output of the generator shown on one of the slices of a case: (a) original image with local maxima of response; (b) after suppression of low-response candidates.}
\label{fig:deletion}
\end{figure}

The fourth and final step of the algorithm involves deletion of the candidates with the normalized LoG filter response value lower than the threshold $R_{CT}=226$, which was determined earlier. An example outcome of the deletion on a selected slice of a CT image is shown in Figure~\ref{fig:deletion}. Here each circle represents a nodule candidate with diameter directly related to the scale at which it was detected.

\subsection {Generation of nonsolid candidates}
\label{nonsolid_section}

Detection of nonsolid nodules is based on an understanding of their unique attenuation characterisitcs in a CT scan. With the exception of the image artifacts caused by heart and respiratory motion, there are no other normal volumetric structures within the lungs that have the same attenuation characteristics as nonsolid lesions. While there are a large number of voxels in the same image intensity range caused by partial volume effects at the edges of pulmonary structures, very few of them are incorporated into objects similar to nodules in size and shape. Therefore, the detection process should identify large regions of blob-like shapes having intermediate image intensity levels between parenchyma and solid tissue inside the lung region. To a first approximation, this is similar to the detection of solid nodules, but with a different target intensity range. 

\begin{figure}
		\centering
		\includegraphics[width=0.75\linewidth]{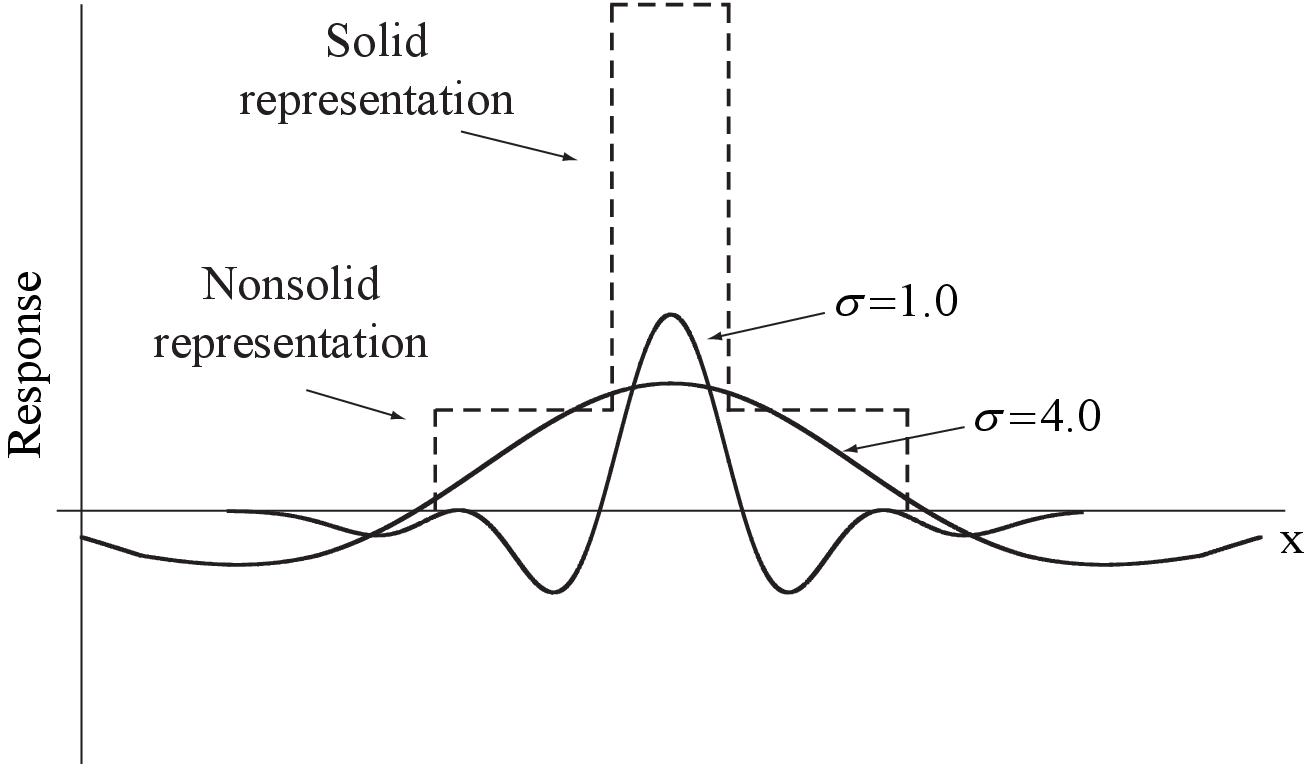}
		\caption{\label{fig:nonsolidinterference} Illustration of the interference effect between solid and nonsolid nodule representations. Response to solid nodule results in higher response causing nonsolid nodule to be missed by the multiscale detector.}
\end{figure}

However, with the presence of pulmonary structures within a nonsolid region, the normalized LoG filter would produce a higher response on such structures, rather than on the entire nonsolid region due to the interference. This is illustrated in example shown Figure~\ref{fig:nonsolidinterference}. Here two nodules are represented by two superimposed rectangular functions of different intensity and size. Responses of the best matching kernels are shown as solid lines. The rectangle of higher intensity resulted in higher response and therefore caused the superimposed low-intensity rectangle to be "missed" by the detector. A similar situation occurs when a nonsolid nodule is located very close to a solid pulmonary vessel.

\begin{figure}
		\centering
		\includegraphics[width=0.75\linewidth]{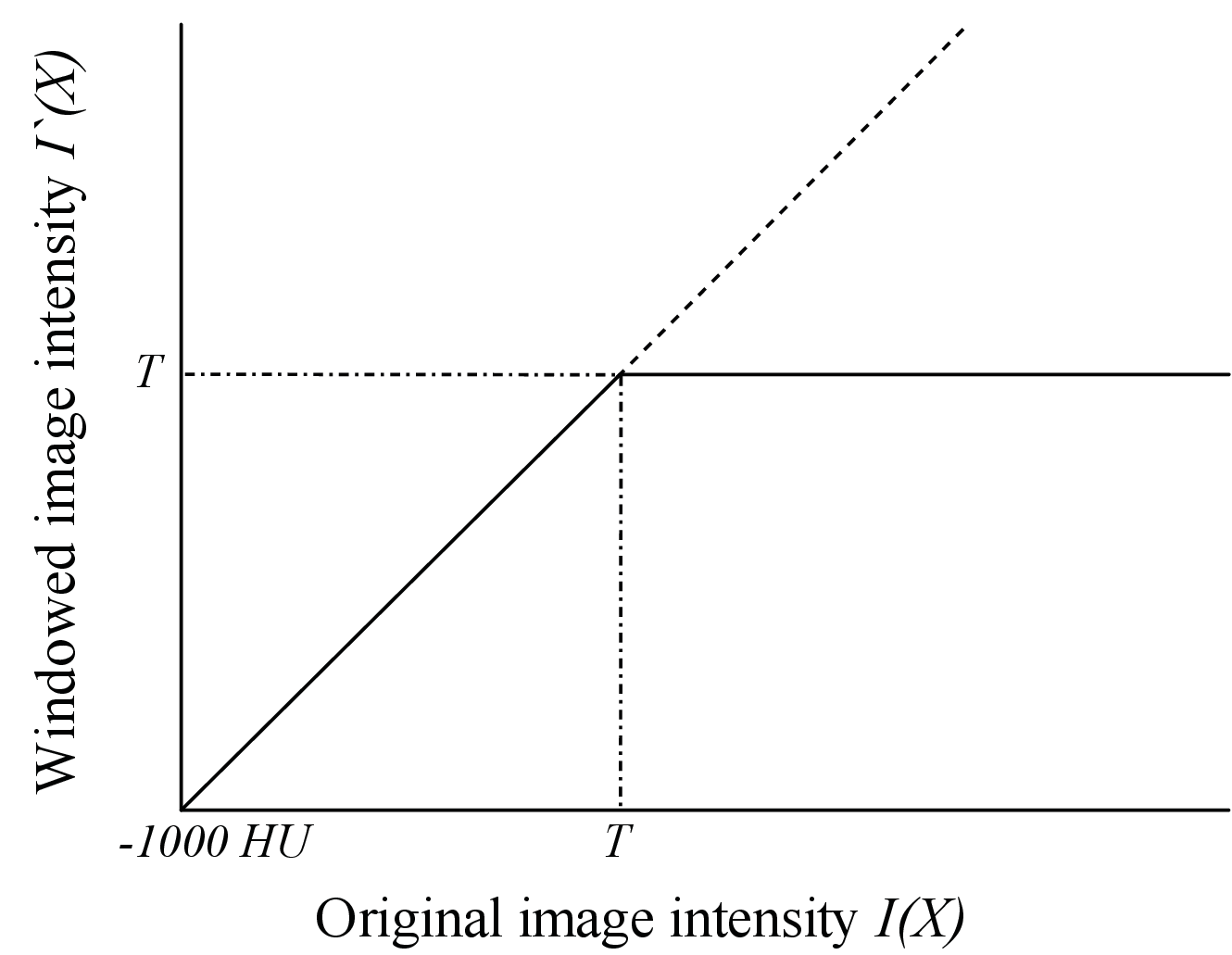}
		\caption{\label{fig:mapping} Image windowing technique. Intensity transform function is applied to the original image to suppress solid components with intensity over $T$.}
\end{figure}

In order to overcome this issue, the image intensity windowing was used. The main purpose of this was to suppress high intensity objects and reduce their interference. The candidate generator was modified for the detection of nonsolid nodules by preprocessing the image with a thresholding filter so that no regions of the image had a higher intensity than $T$. That is, the image was prefiltered as
\begin{equation}
I'(X) = \max \{I(X),T\},
\end{equation}
which is illustrated in Figure~\ref{fig:mapping}.

If we hypothesize that a solid component does not occupy more than half the volume of a nonsolid nodule, the threshold $T$ can be set equal to the median intensity value obtained for nonsolid nodules. This way the interference from solid structures will be eliminated. The value for $T$ was selected as \mbox{-700 HU}, which is 20~HU below the median intensity value obtained by sampling a set of nonsolid nodules intensities~\cite{WBWB07}. 

\begin{figure}
\centering
\subfigure[]{\label{fig:na}\includegraphics[scale=0.5]{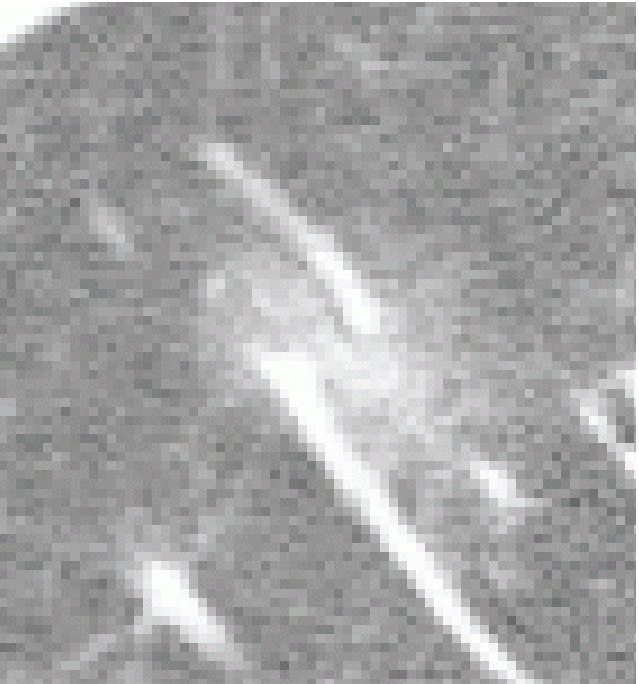}}
\subfigure[]{\label{fig:nb}\includegraphics[scale=0.5]{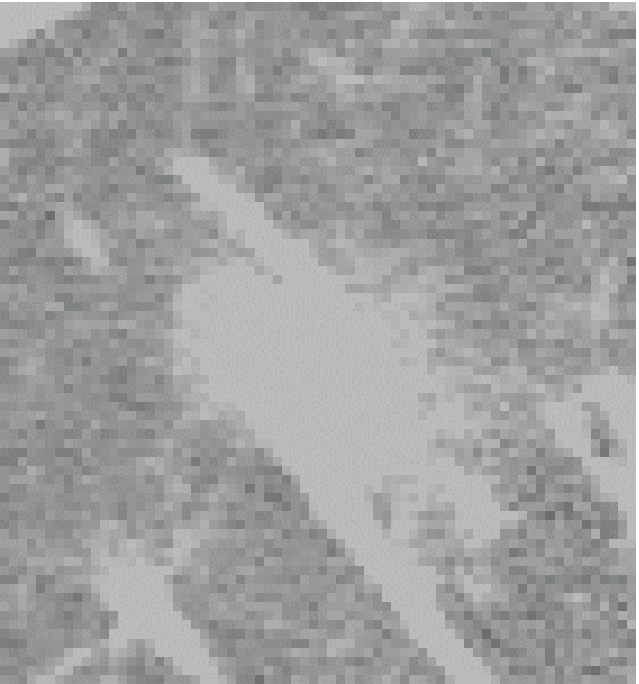}}
\subfigure[]{\label{fig:nc}\includegraphics[scale=0.5]{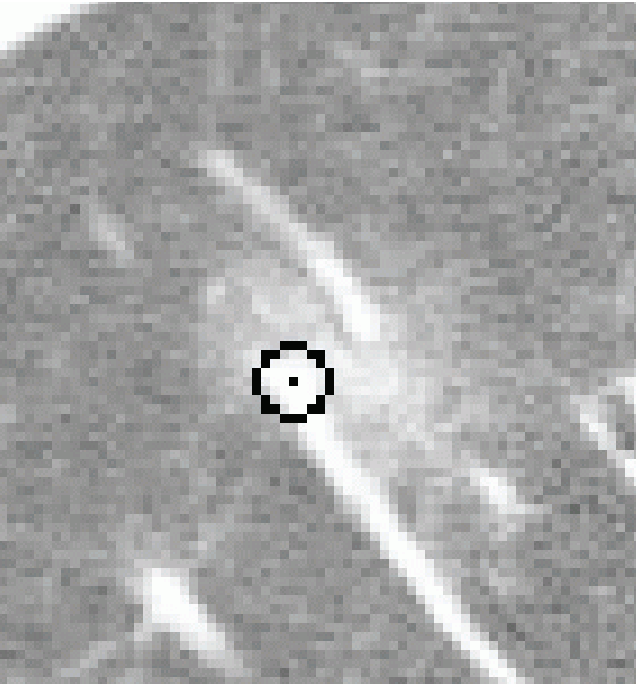}}
\subfigure[]{\label{fig:nd}\includegraphics[scale=0.5]{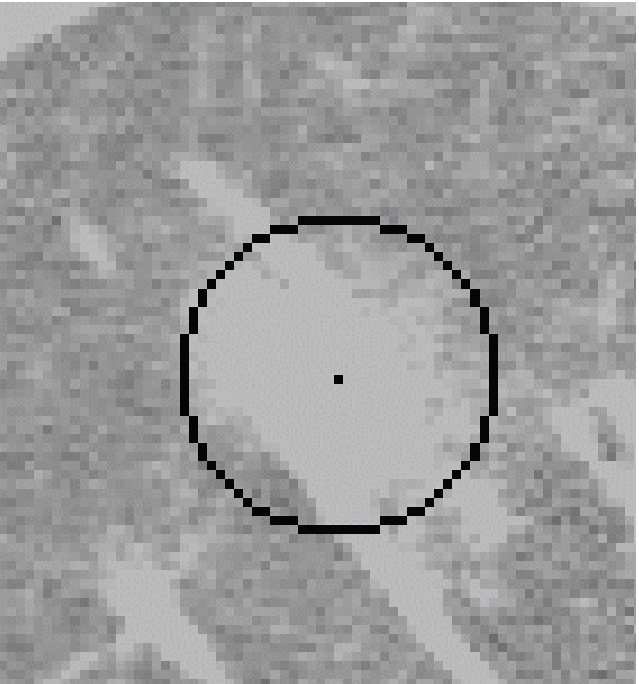}}
\caption{\label{fig:abcd}Result of windowing on a local nodule subregion at a level of \mbox{-680 HU}: (a) original nonsolid nodule image; (b) transformed image; (c) incorrect output of the candidate generator on the original image; (d) correct output of the candidate generator on the transformed image.}
\end{figure}

The illustration of such high intensity suppression on a sample nonsolid region is given in Figure~\ref{fig:abcd}. With the windowing, nonsolid regions should generate a strong response in LoG filtering and provide correctly located and sized nodule candidates.

Therefore, the nonsolid nodule candidate generation consists of a new prefiltering stage followed by the same nodule candidate generation used for detecting solid nodules.

\section {Evaluation of the candidate generator}

The purpose of this section is to show the practical advantage of using normalized LoG filtering scheme for identification of solid and nonsolid pulmonary nodules on a large dataset with respect to the requirements of the candidate generator. These requirements were defined in the introduction section and include: high sensitivity close, accurate size estimation and positional accuracy, and high speed of detection.

To show the effectiveness of the new scheme, a comparison of the normalized LoG-based solid nodule candidate generator to the reference generator was made. The reference generator was previously developed within our research group and described in detail in Enquobahrie et al.~\cite{AEAR07}. The common set of performance measures obtained from the same datasets were obtained for both of the methods.

\subsection{Method}
The candidate generation scheme was evaluated with respect to a nodule size enriched dataset of 706 whole-lung low-dose CT scans 1.25 mm thick from Weill Cornell Medical Center database. All scans in this dataset were reviewed by at least two thoracic radiologists. 
For each identified nodule, one of the experts selected the central slice and measured nodule dimensions corresponding to the nodule's longest transverse axis (length) and its longest perpendicular axis (width). Both axes were not necessary parallel to the image coordinate axes. Length and width measurements were quantized to the step of 0.5 mm. Along with the nodule length and width, approximate location of the nodule centroid expressed as integer pixel coordinates was recorded.

The effective diameter $d(k_i)$ of the nodule $k_i$ from its width $w(k_i)$ and height $h(k_i)$ were determined by:
\begin{equation}
d(k_i) = \frac{1}{2}\left(w(k_i) + h(k_i)\right).
\end{equation}

The evaluation dataset consisted initially of  250 sequential asymptomatic cases from a lung cancer screening study~\cite{IELCAP}, and was enriched with 456 new cases containing at least either one solid nodule with effective diameter greater or equal to 4 mm or one nonsolid nodule greater or equal to 6 mm. With the addition of the enriched data, the fraction of nodules greater than 10 mm increased from 3\% to 6\% and the fraction of nodules greater than 4 mm increased from 23\% to 43\%. 

\begin{figure}
\centering
\subfigure[]{\label{fig:solidhist}\includegraphics[width=0.9\linewidth]{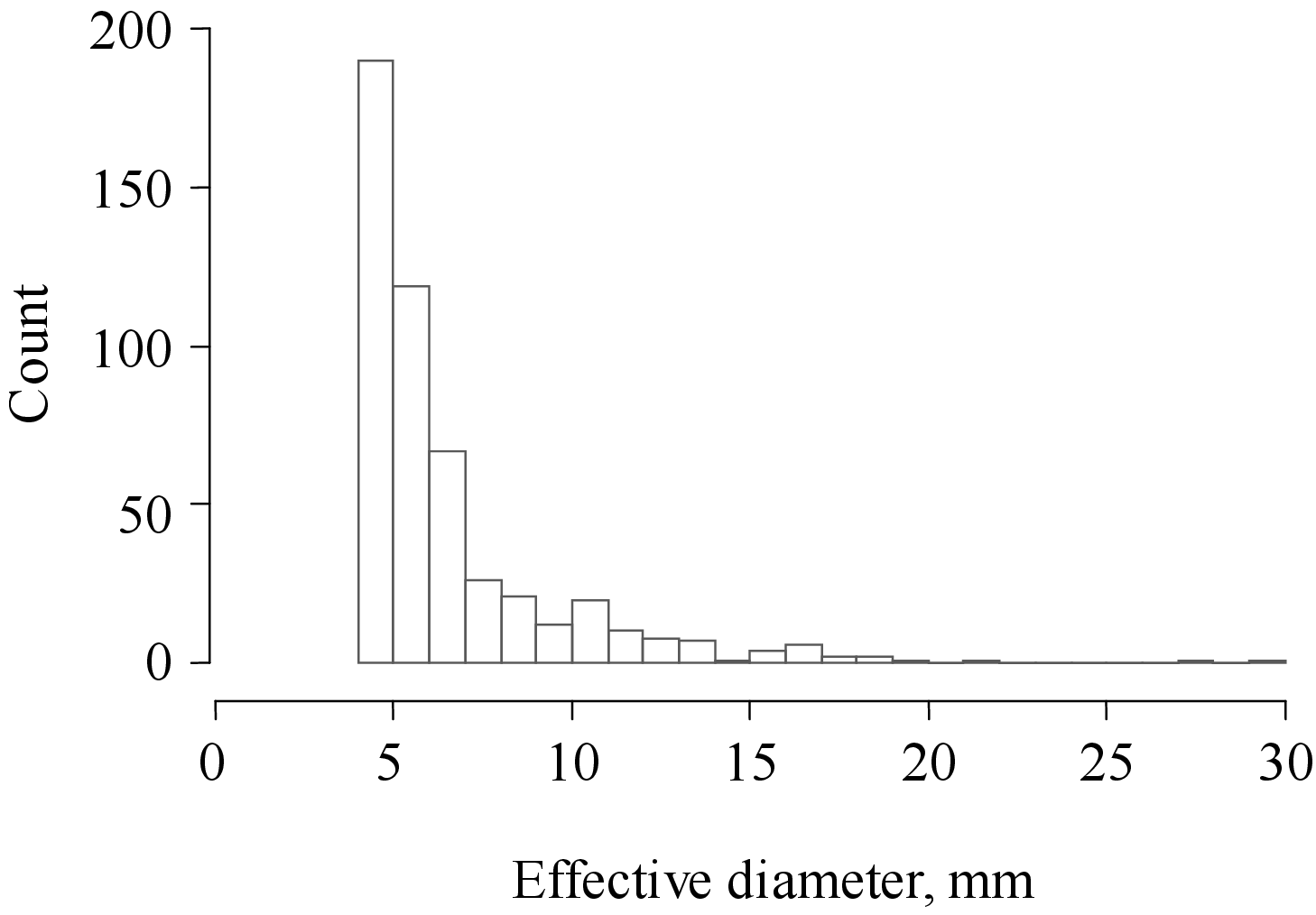}}
\subfigure[]{\label{fig:nonsolidhist}\includegraphics[width=0.9\linewidth]{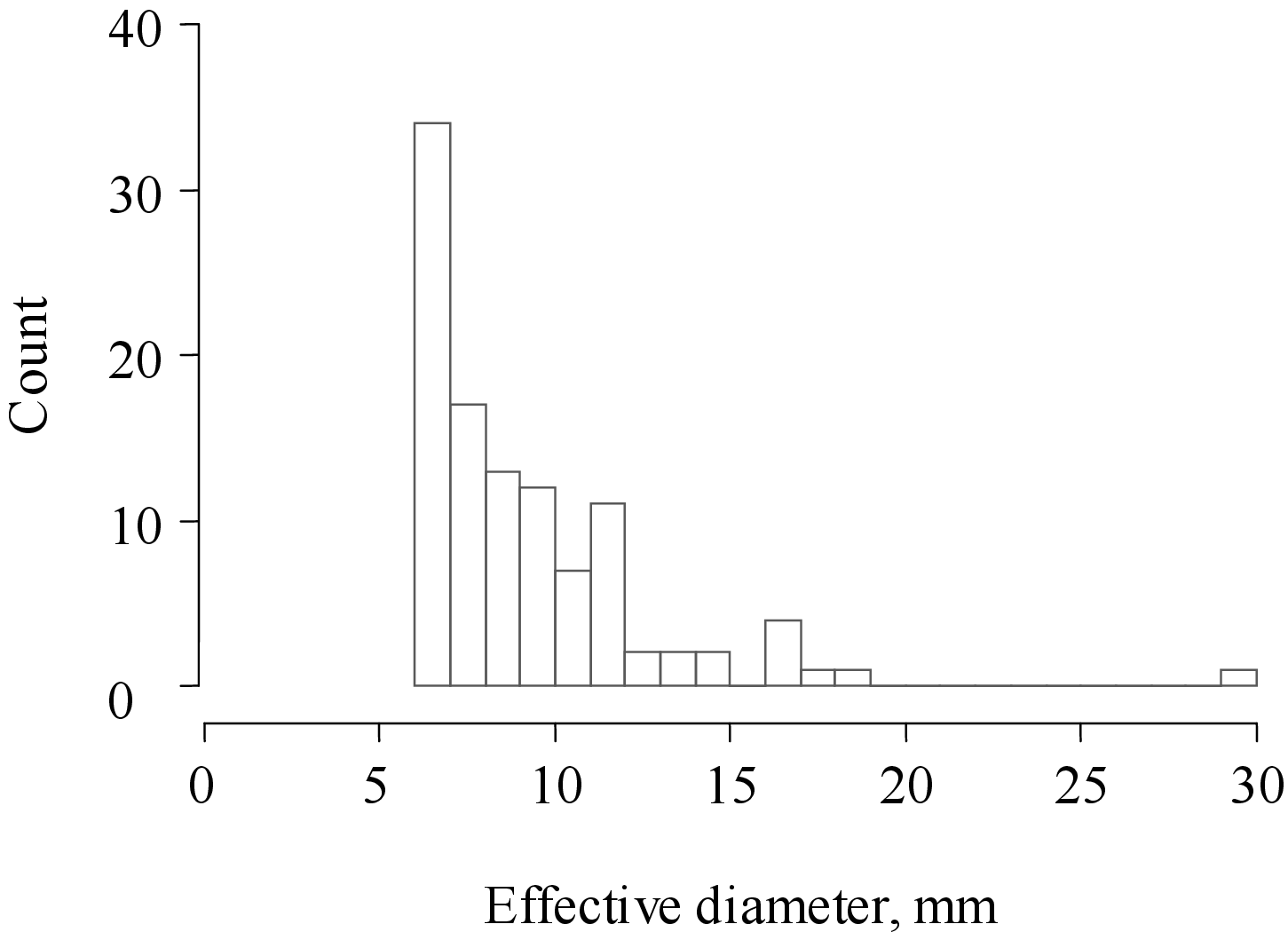}}
\caption{Distributions of solid (a) and nonsolid (b) nodule effective diameters in the evaluation dataset.}
\label{fig:distr}
\end{figure}

Nodules with effective diameter greater or equal to the lower size cut off threshold made up the evaluation set. In the experiment, solid nodules of an effective measured diameter equal or greater than 4 mm (499 nodules) were used. For nonsolid nodules, the diameter cut off was selected to be 6 mm (107 nodules). The value of both size thresholds was slightly lower than the one used in clinical practice to establish the safety margin covering the disagreement between automatic and manual nodule size measurements. The distributions of the nodule sizes for the final dataset are shown in Figure~\ref{fig:distr}. 

The conventional way to assess sensitivity is to measure the fraction of correctly identified nodules over the total number of nodules in the evaluation set. If we denote the set of all nodules that are being targetted as $K$, the final sensitivity $S$ can be calculated as the fraction of correctly identified nodules $n_c$ over the total number of nodules $|K|$. The criterion, confirming that a nodule has been detected, is important and may have an impact on the sensitivity: a nodule $k_i$ was considered as identified, if there existed a corresponding candidate $c_j$, such that the Euclidean distance between their centroid locations was less or equal to the half-length of the nodule $k_i$. The Algorithm~\ref{alg:alg3} shows how the sensitivity is calculated.

\begin{algorithm*}
\def\la{\leftarrow}
\def\FOR{\textbf{for} } 
\def\FOREACH{\textbf{for each} }
\def\TO{to }
\def\STEP{step }
\def\DO{\textbf{do} }
\def\ENDFOR{\textbf{end for} }
\def\IF{\textbf{if} }
\def\THEN{\textbf{then} }
\def\ENDIF{\textbf{end if} }
\def\GOTO{\textbf{go to} }
\def\STOP{\textbf{stop} }
\caption{Calculation of detection sensitivity}
\label{alg:alg3}
\begin{tabular}[t]{l l}
&\\
$K$ & evaluation set of nodules \\
$C$ & set of detected candidates \\
\end{tabular}

\medskip\hrule\medskip

\begin{tabular}[t]{l l}

$n_c \la 0$  & set the number of correctly identified nodules \\
\FOREACH $k_i \in K$  & for each nodule \\
\quad\FOREACH $c_j \in C$  & for each candidate \\
\quad\quad\IF $dist(c_j, k_i) \le 0.5\ l(k_i)$  & if the distance between the centroids is less than the half length of the nodule \\
\quad\quad\quad $n_c \la n_c + 1$  & increment the number of correctly identified nodules \\
\quad\quad\ENDIF  & \\
\quad\ENDFOR  & \\
\ENDFOR  & \\
$S \la n_c / |K|$  & calculate the sensitivity\\

\end{tabular}
\end{algorithm*}

The sensitivity was evaluated independently for solid and nonsolid nodules with the intensity windowing preprocessing. 

The accuracy of the size estimation was evaluated through a comparison of the effective nodule diameter derived from the ground truth to the diameter obtained from the generator. The bias for diameter estimation for the set of detected nodules was calculated as the average difference between the effective diameter $d(k_i)$  of the nodule and the diameter of corresponding candidate as $d(c_j)$:
\begin{equation}
DiameterBias = \mathop E\limits_{c_j \leftrightarrow k_j} \left[ d(k_i) - d(c_j) \right].
\end{equation}

Positional accuracy was evaluated by calculating the average distance between the centroids of the detected nodules and corresponding candidates.

Average candidate generation execution time per CT image was measured by running the candidate generators on Intel Xeon 3.00 GHz processor given that the $Lungs$ mask is already segmented.

\section{Results}

The sensitivity for solid nodules achieved a value of 0.998 (498/499). One solid nodule in the apical region of the lungs was missed due to the inaccuracy of the lung segmentation. Its location was outside the lung mask boundary even after the morphological expansion.

The normalized LoG-based candidate generator considered in this work outperformed the reference generator in identifying solid nodules. The comparison of the generators with respect to the solid nodules of diameters 4 mm and greater is shown in Table~\ref{tab:candresults}.

\begin{table*}
\caption{Comparison of LoG-based and reference generators with respect to detection of solid nodules of diameter of 4 mm and above.} 
\label{tab:candresults}
\begin{center}
\small
\begin{tabular}{|l|l|l|}
\hline
\multicolumn{1}{|c|}{Parameter} & \multicolumn{1}{c|}{LoG-based} & \multicolumn{1}{c|}{Reference} \\ 
\hline
Sensitivity & \multicolumn{1}{c|}{0.998 (498/499)} & \multicolumn{1}{c|}{0.958 (478/499)} \\ 
Diameter estimation accuracy, mm & \multicolumn{1}{c|}{-0.12 $\pm$ 3.27} & \multicolumn{1}{c|}{-1.20 $\pm$ 5.45 } \\ 
Positional accuracy, mm  & \multicolumn{1}{c|}{1.40} & \multicolumn{1}{c|}{1.66} \\ 
Size range of detected candidates, mm & \multicolumn{1}{c|}{3.00 -- 25.00} & \multicolumn{1}{c|}{1.96 -- 22.34} \\ 
Average number of candidates per case & \multicolumn{1}{c|}{8177} & \multicolumn{1}{c|}{3785} \\ 
Average CT image processing time, minutes & \multicolumn{1}{c|}{4.5} & \multicolumn{1}{c|}{8.0} \\ 
\hline
\end{tabular}
\end{center}
\end{table*}

Sensitivity of the reference generator on the evaluation set was lower. Its scheme for detection of attached nodules was highly dependent on the correctness of the lung segmentation, therefore, the majority of the false negatives were located on periphery of the lung. Even though both LoG-based and reference generators resulted in an overestimation bias for nodule diameter, the LoG-based generator was substantially more accurate and resulted in smaller error: the 95\% confidence intervals of the difference between automated and manual nodule size measurements were \mbox{-0.12 $\pm$ 3.27} mm for solid nodules.  

The normalized LoG-based candidate generator also turned out to be more accurate than the reference generator in centroid estimation as its average nodule candidate distance of 1.40 mm was lower than 1.66 mm. This could be partially explained by the fact that the reference algorithm locates the peaks for attached nodules as opposed to their centroids.

For nonsolid nodules, the LoG-based candidate generator achieved the perfect sensitivity value of 1.000 (107/107). 


The generator of nonsolid candidates resulted in a size overestimation of -1.27 $\pm$ 5.70 mm, while the average candidate location accuracy was 1.43 mm for nonsolid nodules. 

Each solid and nonsolid candidate generation algorithms were able to complete the processing of a single scan in 4.5 minutes on average.

\section{Discussion}

Both solid and nonsolid candidate generators achieved very high detection sensitivities of 99.8\% and 100.0\%, which makes the detection systems based on the normalized LoG filter a promising solution for the detection of pulmonary nodules. Candidate generation with windowing prefiltering has been found to be a powerful technique.

LoG-based candidate generation resulted in size overestimates for both solid and nonsolid nodules. One of the possible reasons for this is that the multiscale LoG response function is asymmetric~(\ref{eqn:response}). The overestimate was greater for nonsolid nodules, probably, because of greater average size, and the subjectivity in size measurements by radiologists: the boundary of a nonsolid nodule is not clearly defined and therefore the bias in candidate size estimation may also depend on how the operator set up the windowing level.

The experiments have also shown that nonsolid nodules could be identified with almost the same positional accuracy (1.40 mm vs. 1.43 mm) as solid nodules. 

\begin{figure}
		\centering
		\includegraphics[width=0.85\linewidth]{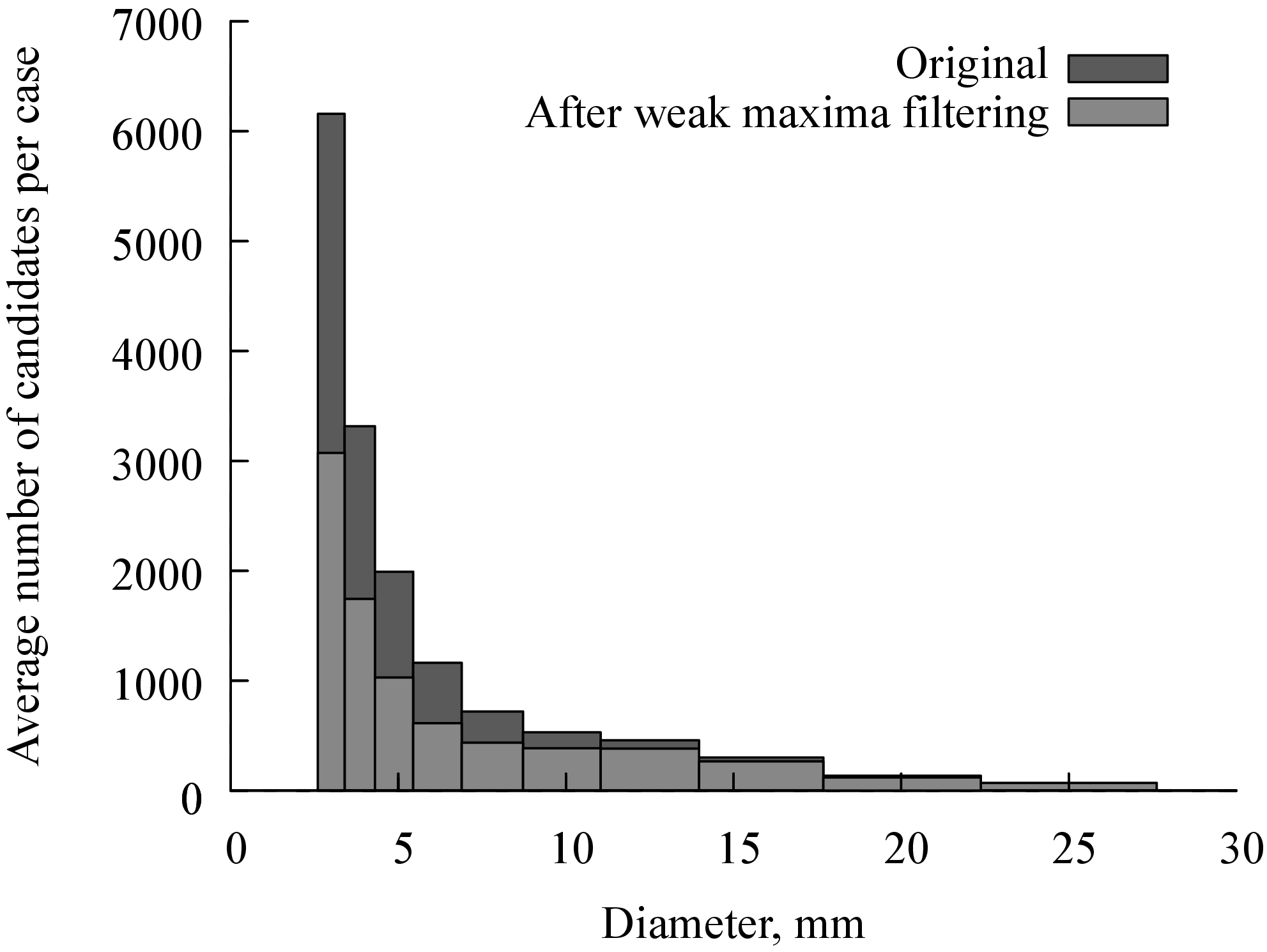}
		\caption{\label{fig:canddistr} Distribution of candidate sizes before and after suppression of low response candidates. Each bin corresponds to one of 10 discrete diameter levels used by the generator.}
\end{figure}

The distribution of solid candidate responses for the set prior to deleting candidates with low filter response is shown in Figure~\ref{fig:threepeaks}. The graph has three local modes that were visually identified as corresponding to: (a) small intensity variations (noise) in lung parenchyma; (b) lung and mediastinal surface irregularities including some attached nodules; (c) pulmonary vessels (including branch points), airways, tips of the ribs, calcification and remaining nodules. The lowest normalized response for a solid nodule was 228, as compared to the theoretically computed value of 226 which means all the true candidates above the threshold were preserved. The result of rejecting low-response candidates is illustrated in Figure~\ref{fig:canddistr}. The plot shows the distribution of candidate sizes with respect to their estimated size. From this graph one can infer the expected number of candidates generated for a given size range. The number of candidates increases rapidly with the decrease of candidate size and therefore, the generator can be tuned to a specific target nodule size range. 
\begin{figure}
		\centering
		\includegraphics[width=0.85\linewidth]{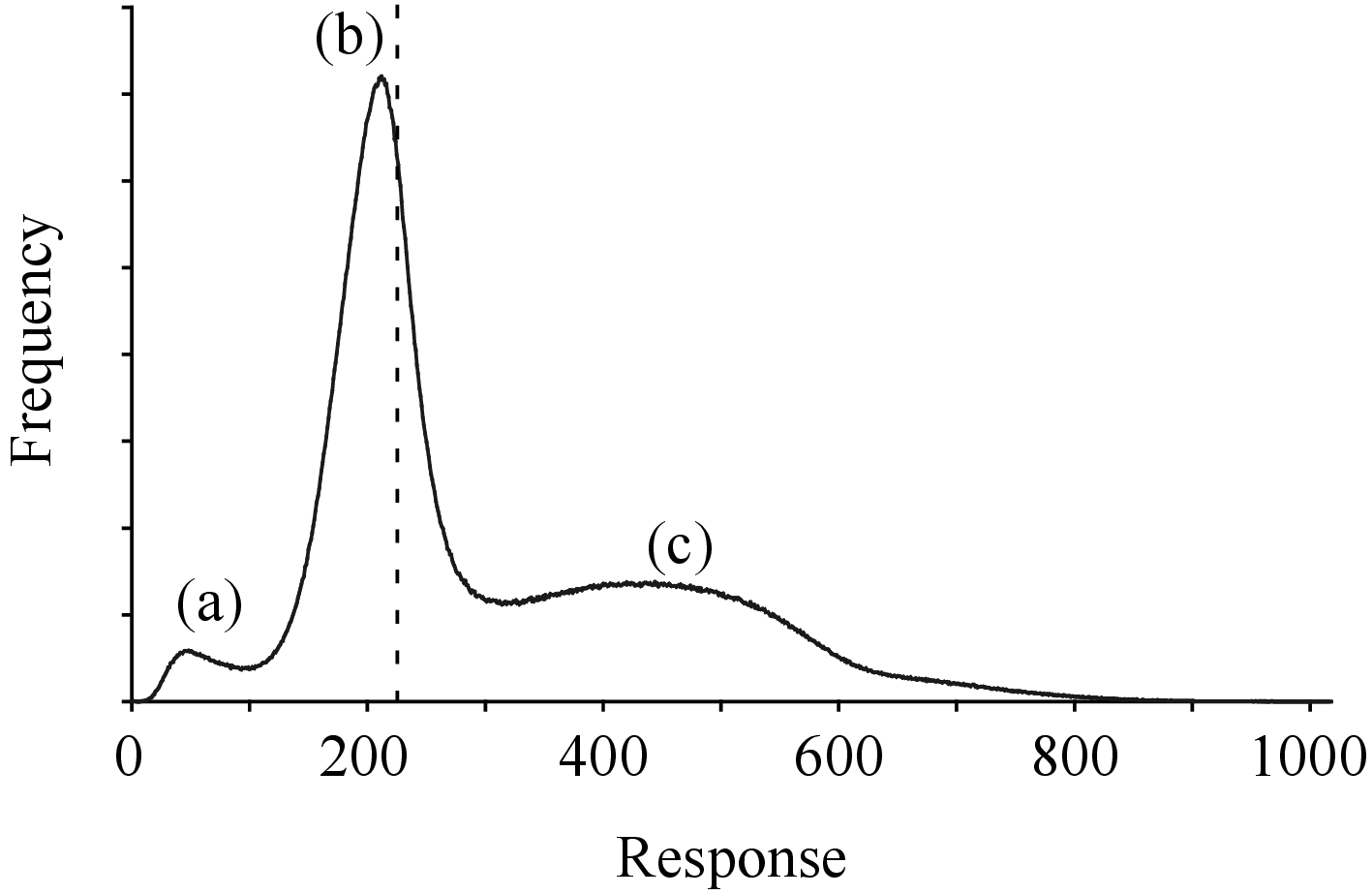}
		\caption{\label{fig:threepeaksdistr} Distribution of the candidate normalized responses shows three distinct peaks corresponding to: (a) noise in lung parenchyma, (b) lung surface irregularities including attached nodules and (c) pulmonary vessels (including branch points), airways and remaining nodules with lesser degree of attachments. Dashed line is the response threshold.}
\end{figure}

The speed of the presented candidate generator is mostly limited by the time needed for calculating discrete Fourier transforms and depends on the image pixel dimesions of the image and number of scale discretization levels. Further optimization either by downsampling of the original CT scan or by employing more efficient filtering algorithms, such as suggested by Jin and Gao~\cite{jin1997recursive} may decrease the computation time.

The proposed candidate generator achieves higher sensitivity but at a cost of a substantial increase in false positives (this is also the case for our reference method) when compared to other candidate generators reported in the literature (Table~\ref{tab:perfcomparison}). All generated candidates are typical for a nodule detection system and can be eliminated using standard false positive reduction approaches. For example, elimination of the candidates caused by vessels and surface irregularities, can be done with simple filters that impose shape and density constraints on the candidates~\cite{AEAR07}. Therefore, the burden of higher false positive rate should not be a major impediment in the implementation of the false positive reduction scheme. At the same time, increased sensitivity of the generator sets higher upper bound for the final sensitivity of a CAD system.

\section {Conclusion}

We have presented a CAD design model for which the candidate generator with 100\% sensitivity is considered as a speed enhancement component rather than a nodule-classifying component of a CAD system. In this context, we can consider the performance of the candidate generator without confounding with the FPE performance. We have presented and evaluated a candidate generator design based on the LoG filtering that meets the high sensitivity requirement with a reasonable number of false positives. We have also demonstrated that, if necessary, the outcome of our candidate generator can be modified such that a similar performance false positive rate to other reported candidate generator designs can be accomplished with some small loss of sensitivity. However, the advantage of the proposed design model is that it facilitates the development of CAD systems with a higher sensitivity than more traditional approaches.

\appendices
\section{Determining normalized LoG scale for detecting one-dimensional rectangular profile}
\label{apx0}
We need to determine parameter $\sigma$ of one-dimensional LoG kernel resulting in a highest response when imposed on a one-dimensional rectangular function of width $d$, centered at the origin and defined as
\begin{equation}
	B(x, d) = \begin{cases} 1, & \text{if $|x| \le d/2$}, \\
											 0, & \text{otherwise}.
\end{cases}
\end{equation}
The value of the response function at $x = 0$, the center of the rectangular function, is equal to
\begin{equation}
\begin{split}
R(\sigma, d) = & \int\limits_{-\infty}^{+\infty} B(x, d) \nabla_{norm}^2G(x,\sigma)\,dx =\\
          = & \int\limits_{-d/2}^{+d/2} -\sigma^2 \nabla^2 \left(\frac{1}{\sqrt{2\pi}\sigma}e^{-\frac{x^2}{2\sigma^2}}\right) \,dx = \\
          = & \left.\frac{x}{\sqrt{2\pi}\sigma}e^{-\frac{x^2}{2\sigma^2}}\right|_{-d/2}^{+d/2}
          = \frac{d}{\sqrt{2\pi}\sigma}e^{-\frac{d^2}{8\sigma^2}}.
\end{split}
\end{equation}
Maximum of the response function can be determined by finding derivative with respect to $d$
\begin{equation}
\frac{\partial R(\sigma, d)}{\partial \sigma} = \frac{d e^{-\frac{d^2}{8\sigma^2}}}{\sqrt{32\pi}\sigma^2} \left(\frac{d^2}{4\sigma^2} - 1\right)
\end{equation}
and solving $\partial R(\sigma, d)/ \partial \sigma = 0$ with respect to $\sigma$. After rejecting negative roots, we obtain the optimal value $\sigma = d / 2$.

\section{Optimization of the convolution computation}
\label{apx1}
One efficient implementation of the multiscale LoG transform is to do the computation of the convolutions in the frequency domain. 

The original expression for response function for the scale $\sigma_i$
\begin{equation}
\nabla_{norm}^2L(X,\sigma_i) =-\sigma_i^2\nabla^2G(X,\sigma_i)*I(X)
\end{equation}
can be rewritten using the convolution theorem as 
\begin{equation}
\begin{split}
&\nabla_{norm}^2L(X,\sigma_i) = \\
&= -\sigma_i^2 \cdot \mathcal{F}^{-1}\left\{\mathcal{F}\{\nabla^2G(X,\sigma_i)\}\cdot \mathcal{F}\{I(X)\}\right\},
\end{split}
\end{equation}
where $\mathcal{F}$ and $\mathcal{F}^{-1}$ denote forward and inverse Fourier transforms.

Fourier transforms of the LoG is known and can be precomputed in advance:
\begin{equation}
\begin{split}
&\mathcal{F}\left\{\nabla^2G(X,\sigma_i)\right\} = \\
&= -{(2\pi)^{-1.5}}{\Omega^T}\Omega \exp \left(-0.5\sigma_i^2\Omega ^T\Omega\right),
\end{split}
\end{equation}
where $\Omega$ is a three-dimensional vector of angular frequencies $\Omega=(\omega_x,\omega_y,\omega_z)$.

This way, in order to compute multiple convolutions for an image, one needs to take the Fourier transform once, and for each scale multiply it with precomputed transform of the LoG. Taking the inverse transform on the result and multiplying it by normalization coefficient for each scale would result in desired response function.

\section*{Acknowledgment}

This research was supported in part by NIH grant R33CA101110 and the Flight Attendants' Medical Research Institute.




\bibliographystyle{IEEEtran}
\bibliography{paper}
\end{document}